\documentclass[a4paper,11pt]{article}
\pdfoutput=1
\usepackage{jcappub}
\usepackage{soul}
\usepackage{appendix}

\newcommand{\mdm}{m_\text{DM}}
\newcommand{\Tbh}{T_\text{BH}}
\newcommand{\Tin}{T_\text{BH}^\text{in}}
\newcommand{\Tev}{T_\text{evap}}
\newcommand{\Teq}{T_\text{eq}}
\newcommand{\Min}{m_\text{in}}
\newcommand{\Trh}{T_\text{rh}}
\newcommand{\Tmax}{T_\text{max}}

\usepackage[normalem]{ulem}
\DeclareUnicodeCharacter{3B1}{\ensuremath{\alpha}}

\begin{document}
\title{Cogenesis of Baryon Asymmetry and Gravitational Dark Matter from Primordial Black Holes}
\author[a]{Basabendu Barman,}
\author[b]{Debasish Borah,}
\author[b]{Suruj Jyoti Das,}
\author[c]{and Rishav Roshan}

\affiliation[a]{\,Centro de Investigaciones, Universidad Antonio Nari\~{n}o\\Carrera 3 este \# 47A-15, Bogot{\'a}, Colombia}
\affiliation[b]{\,Department of Physics, Indian Institute of Technology Guwahati, Assam 781039, India}
\affiliation[c]{Department of Physics, Kyungpook National University, Daegu 41566, Korea}
\emailAdd{basabendu88barman@gmail.com}
\emailAdd{dborah@iitg.ac.in}
\emailAdd{suruj@iitg.ac.in}
\emailAdd{rishav.roshan@gmail.com}

\abstract{We propose a scenario where dark matter (DM) with a wide mass range from a few keV to PeV can be produced solely from evaporating primordial black holes (PBH), while being consistent with the required free streaming length for structure formation. If DM does not have any other interactions apart from gravity and the universe has a PBH dominated phase at early epoch, then PBH evaporation typically leads to overproduction of DM in this mass range. By incorporating this gravitational DM within a Type-I seesaw scenario with three right handed neutrinos (RHN), we bring the abundance of PBH generated DM within observed limits by late entropy injection due to decay of one of the RHNs, acting as the diluter. The diluter, due to its feeble coupling with the bath particles, gets produced primarily from the PBH evaporation thereby leading to the second stage of early matter domination after the end of PBH dominated era. The other two RHNs contribute to the origin of light neutrino mass and also lead to the observed baryon asymmetry via leptogenesis with contributions from both thermally and PBH generated RHNs. The criteria of DM relic and baryon asymmetry can be satisfied simultaneously if DM mass gets restricted to a ballpark in the MeV-GeV regime with the requirement of resonant leptogenesis for heavier DM mass in order to survive the large entropy dilution at late epochs.}

\begin{flushright}
  PI/UAN-2022-714FT \\
\end{flushright}
\maketitle


\section{Introduction}
\label{sec:intro}
Observational evidences from astrophysics and cosmology based experiments suggest that we live in a baryon asymmetric universe whose matter content is dominated by a non-luminous, non-baryonic form of matter known as dark matter (DM) \cite{Zyla:2020zbs, Aghanim:2018eyx}. The observed baryon asymmetry of the universe (BAU) is quantitatively quoted in terms of the ratio of excess baryons over antibaryons and photons \cite{Aghanim:2018eyx} 
\begin{equation}
\eta_B = \frac{n_{B}-n_{\overline{B}}}{n_{\gamma}} \simeq 6.2 \times 10^{-10}.
\label{etaBobs}
\end{equation}
Such an asymmetry needs to be generated dynamically as any pre-existing asymmetry is likely to be wiped out by the exponential expanding phase during cosmic inflation. However, the Standard Model (SM) fails to satisfy the criteria, known as Sakharov's conditions~\cite{Sakharov:1967dj}, for dynamical generation of BAU requiring the presence of beyond standard model (BSM) physics. On the other hand, the particle nature of DM is not yet known with none of the SM particles being fit to be DM candidate. Among different BSM proposals for generating BAU, baryogenesis via leptogenesis \cite{Fukugita:1986hr} is one of the most popular scenarios. In such a scenario, a non-zero asymmetry is first created in the lepton sector, and then gets converted into baryon asymmetry via $(B+L)$-violating electroweak sphaleron transitions~\cite{Kuzmin:1985mm}. Interestingly, most leptogenesis framework can also accommodate non-zero neutrino mass and mixings \cite{Zyla:2020zbs}, another observed phenomena which the SM fails to address. Similarly, among different BSM proposals for DM, the weakly interacting massive particle (WIMP) paradigm is one of the most well studied scenarios. In WIMP framework, a particle having mass and interactions around the electroweak ballpark can lead to the observed DM abundance after undergoing thermal freeze-out in the early universe \cite{Kolb:1990vq}. 

While leptogenesis remains a high scale phenomena with limited observational scopes\footnote{Probing high scale leptogenesis via stochastic gravitational wave observations is possible in certain scenarios, as pointed out in \cite{Dror:2019syi}.}, WIMP DM has promising direct detection prospects due to sizeable interactions among DM and SM particles. However, no such signatures have been found in direct detection experiments so far, motivating the particle physics community to explore other viable alternatives. It is also important to note that all the observational evidences supporting the presence of DM in the universe are based on purely gravitational interactions only. Thus, it is appealing to consider the possibility of DM production in the early universe via a mechanism relying on gravitational interactions only \cite{Ford:1986sy}. DM production via minimal coupling to gravity has been discussed in several works  \cite{Chung:1998ua, Chung:1998zb, Kuzmin:1998uv, Kuzmin:1998kk, Greene:1997ge, Chung:1998rq, Garny:2015sjg, Ahmed:2020fhc, Kolb:2020fwh}. Similarly, non-minimal coupling to gravity can also lead to DM production, as discussed already in several works including \cite{Alonso-Alvarez:2018tus, Alonso-Alvarez:2019ixv, Markkanen:2015xuw, Fairbairn:2018bsw, Ema:2016hlw, Ema:2018ucl, Cembranos:2019qlm, Chung:2018ayg, Babichev:2020xeg, Borah:2020ljr, Barman:2021qds}. In this work, we consider another alternative: production of gravitational DM from evaporation of primordial black holes (PBH) assuming the latter to dominate the energy density of the universe at some stage. Contrary to earlier works where either light or superheavy DM production from evaporating PBH was shown, here we show the possibility of a much wider range allowing the intermediate mass DM: from a few keV to PeV. This is achieved by considering the presence of three heavy right handed neutrinos (RHN) which also takes part in Type-I seesaw mechanism \cite{Mohapatra:1979ia,Yanagida:1979as,GellMann:1980vs,Glashow:1979nm} of neutrino mass generation. DM with mass in the intermediate regime mentioned above usually gets overproduced from PBH evaporation and can be brought within Planck limits by late entropy injection from one of the right handed neutrinos which is long-lived and couples feebly to SM neutrinos resulting in vanishingly small lightest active neutrino mass. The other two RHNs can not only reproduce neutrino oscillation data via Type-I seesaw but also can produce the baryon asymmetry of the universe via leptogenesis.

This paper is organised as follows. In section \ref{sec:dm-pbh}, we briefly discuss our basic setup consisting of three right handed neutrinos, a particle DM candidate with only gravitational interactions and a PBH dominated phase. In section \ref{sec:pbh}, we provide some analytical estimates for PBH generated DM and leptogenesis from PBH generated RHN decay. In section \ref{sec:numerical}, we discuss the details of our numerical analysis followed by brief discussion on production of DM and RHN from gravity mediated scatterings in section \ref{sec:grav}. Finally, we conclude in section \ref{sec:concl}.

\section{The Basic Framework}
\label{sec:dm-pbh}
As mentioned before, we consider a simple BSM framework where the SM is extended by three generations of right handed neutrinos $N_i\,(i=1,2,3)$ and a dark matter particle, all of which are singlets under the SM gauge symmetry. 
Although
three copies of RHNs are considered in typical Type-I seesaw model, two are sufficient to fit light neutrino data. However, we require three copies to realise the cogenesis of DM and baryon asymmetry as we discuss in upcoming sections. The interaction Lagrangian reads
\begin{align}
& -\mathcal{L}\supset\frac{1}{2}\,M_N\,\overline{N^c}N+y_N\,\overline{N}\,\tilde{H}^{\dagger}\,\ell +{\rm h.c.}\,,     
\end{align}
where we have suppressed the generation index and the RHNs are supposed to be mass diagonal. The SM leptons are denoted by $\ell$, while $\tilde{H}\equiv i\,\sigma_2\,H^\star$, $H$ being the SM Higgs and $\sigma_i$ are the Pauli spin matrices. The Dirac Yukawa coupling of neutrinos get fixed via Casas-Ibarra parametrisation (see Appendix.~\ref{sec:app-CI}), after using the best-fit values of light neutrino parameters \cite{Zyla:2020zbs}. Since DM is assumed to have only gravitational interactions, it only has a mass term, the exact form of which depends on the particle nature of DM particle. In our numerical analysis, for simplicity, we will consider a real singlet scalar $S$ of mass $m_\text{DM}$, to be a potential DM candidate\footnote{We assume the strength of the portal interaction $S^2\,\left|H\right|^2$ is absent by construction to ensure only gravitational production of DM. Considering DM with a different spin does not alter the outcome of our analysis.} which is produced purely from PBH evaporation. One of the RHNs, $N_3$ in the present scenario, is considered to be long lived such that its decay rate is suppressed compared to the other two RHNs. As a consequence, the universe undergoes a $N_3$-dominated epoch even after PBH is completely evaporated. The other two RHNs, namely $N_{1,2}$ can undergo CP-violating out of equilibrium decay, producing lepton asymmetry which can then be converted into the baryon asymmetry of the universe, thanks to the electroweak sphaleron transition. Sphaleron interactions are in equilibrium in the temperature range between $\sim 100$ GeV and $10^{12}$ GeV, and they convert a fraction of a non-zero $B-L$ asymmetry into a baryon asymmetry via
\begin{align}
& Y_B\simeq a_\text{sph}\,Y_{B-L}=\frac{8\,N_F+4\,N_H}{22\,N_F+13\,N_H}\,Y_{B-L}\,,    
\end{align}
where $N_F$ is the number of fermion generations and $N_H$ is the number of Higgs doublets, which in our case $N_F=3\,,N_H=1$ and $a_\text{sph}$ turns out to be 28/79. In leptogenesis, where purely a lepton asymmetry is generated, $B-L=-L$. This is converted into the baryon asymmetry via sphaleron transition~\cite{Buchmuller:2004nz,Buchmuller:2005eh}. Finally, the observed baryon asymmetry of the universe is given by~\cite{Planck:2018jri}
\begin{align}
\eta_B=\frac{n_B-n_{\overline{B}}}{n_\gamma}\simeq 6.2\times 10^{-10}\,,Y_B\simeq 8.7\times 10^{-11}\,.\label{eq:etab}    
\end{align}
\begin{figure}[htb!]
    \centering
    \includegraphics[scale=0.13]{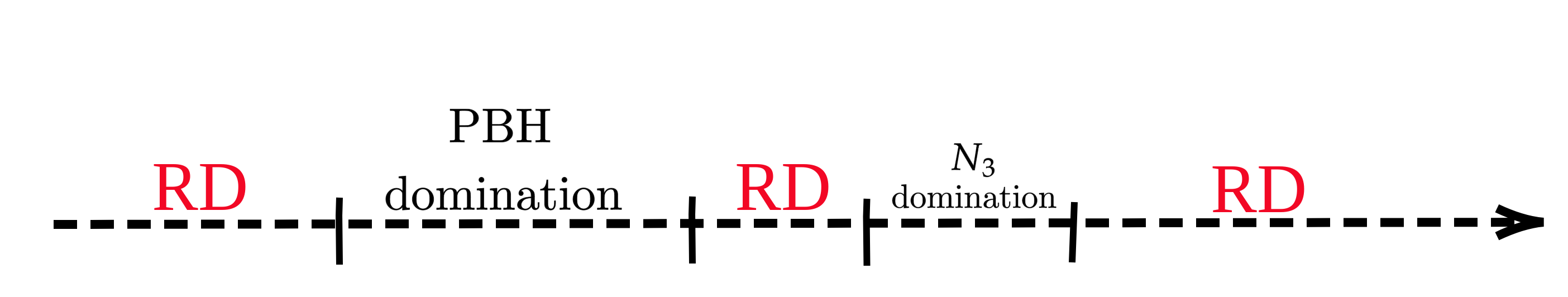}
    \caption{Dominant component of the energy density of the universe at different epoch (time runs from left to right).}
    \label{fig:scheme}
\end{figure}
During the decay of $N_3$ entropy is injected into the radiation bath \cite{PhysRevD.31.681} leading to dilution of the DM and baryon number density produced from PBH. Role of such late entropy injection in tackling DM over-abundance has been studied by several authors in different contexts \cite{Nemevsek:2012cd, Bezrukov:2009th, Borah:2017hgt, Cirelli:2018iax, Dror:2020jzy,Dutra:2021lto, Arcadi:2021doo, Borah:2021inn, Borah:2022byb}. In Fig.~\ref{fig:scheme} we schematically show the components that dominate the energy density of the universe at different epoch. As the time flows from the left to the right, hence, as expected, the Universe is radiation dominated (RD) at the end of reheating. Once the PBH formation happens, the PBH energy density can dominate, which ends as the PBH evanescence is complete. The universe is then dominated by RHN produced from the PBH. Finally, as the RHN decays away, the usual radiation domination era again commences. Note that, in addition to DM, the diluter $N_3$ is also dominantly produced from PBH evaporation as its coupling with SM leptons is tiny due to the requirement of its long-lifetime.

\section{PBH and Cogenesis}
\label{sec:pbh}
Primordial black holes, originally proposed by Hawking \cite{Hawking:1974rv, Hawking:1975vcx}, can have very interesting cosmological signatures \cite{Chapline:1975ojl, Carr:1976zz} (a recent review of PBH may be found in \cite{Carr:2020gox}). While PBH with suitable mass range can itself be a DM candidate, we are interested in its ultra light mass regime where it is not long-lived enough to be DM but can play non-trivial role in production of DM as well as baryon asymmetry. Role of PBH evaporation on DM genesis has been studied in several works \cite{Morrison:2018xla, Gondolo:2020uqv, Bernal:2020bjf, Green:1999yh, Khlopov:2004tn, Dai:2009hx, Allahverdi:2017sks, Lennon:2017tqq, Hooper:2019gtx, Chaudhuri:2020wjo, Masina:2020xhk, Baldes:2020nuv, Bernal:2020ili, Bernal:2020kse, Lacki:2010zf, Boucenna:2017ghj, Adamek:2019gns, Carr:2020mqm, Masina:2021zpu, Bernal:2021bbv, Bernal:2021yyb, Samanta:2021mdm, Sandick:2021gew, Cheek:2021cfe, Cheek:2021odj}. Similarly, the role of PBH evaporation on baryogenesis was first pointed out in \cite{Hawking:1974rv, Carr:1976zz} and has been studied subsequently by several authors in different contexts~\cite{Baumann:2007yr, Hook:2014mla, Fujita:2014hha, Hamada:2016jnq, Morrison:2018xla, Hooper:2020otu, Perez-Gonzalez:2020vnz, Datta:2020bht, JyotiDas:2021shi, Smyth:2021lkn, Barman:2021ost, Bernal:2022pue, Ambrosone:2021lsx}. In a few related works, the possibility of both DM and RHN production from PBH evaporation was considered \cite{Fujita:2014hha, Morrison:2018xla, Hooper:2019gtx, Lunardini:2019zob, Masina:2020xhk, Hooper:2020otu, Datta:2020bht, JyotiDas:2021shi, Schiavone:2021imu, Bernal:2021yyb, Bernal:2021bbv, Bernal:2022swt}. In the present work, we consider DM production solely from PBH evaporation and prevent it from being overproduced by late entropy dilution from one of the RHNs while the other two RHNs produce the baryon asymmetry via leptogenesis. Before proceeding to the detailed numerical analysis, we first show the key features of DM and RHN production from PBH evaporation to set the motivation clear.

We assume PBHs to have formed after inflation during the era of radiation domination. Assuming radiation domination, the mass of the black hole from gravitational collapse is typically close to the value enclosed by the post-inflation particle horizon and is given by~\cite{Fujita:2014hha,Masina:2020xhk}
\begin{equation}
m_\text{BH}^{\rm in}=\frac{4}{3}\,\pi\,\gamma\,\Bigl(\frac{1}{\mathcal{H}\left(T_\text{in}\right)}\Bigr)^3\,\rho_\text{rad}\left(T_\text{in}\right)\,
\label{eq:pbh-mass}
\end{equation}
\noindent with 
\begin{equation}
\rho_\text{rad}\left(T_\text{in}\right)=\frac{3}{8\,\pi}\,\mathcal{H}\left(T_\text{in}\right)^2\,M_\text{pl}^2\,
\end{equation}
where $\mathcal{H}$ is the Hubble parameter, $M_\text{pl}$ is the Planck mass and $\gamma\simeq 0.2$ is a numerical factor which contains the uncertainty of the PBH formation.  As mentioned earlier,  PBHs are produced during the radiation dominated epoch, when the SM plasma has a temperature $T=T_\text{in}$ which is given by
\begin{equation}
T_\text{in}=\Biggl(\frac{45\,\gamma^2}{16\,\pi^3\,g_\star\left(T_\text{in}\right)}\Biggr)^{1/4}\,\sqrt{\frac{M_\text{pl}}{m_\text{BH}(T_\text{in})}}\,M_\text{pl}\,.
\label{eq:pbh-in}
\end{equation}
Once formed, PBH can evaporate by emitting Hawking radiation \cite{Hawking:1974rv, Hawking:1975vcx}. A PBH can evaporate efficiently into particles lighter than its instantaneous temperature $T_\text{BH}$ defined as \cite{Hawking:1975vcx}
\begin{equation}
T_{\rm BH}=\frac{1}{8\pi\,G\, m_{\rm BH}}\approx 1.06~\left(\frac{10^{13}\; {\rm g}}{m_{\rm BH}}\right)~{\rm GeV}\,,
\end{equation}
\noindent where $G$ is the universal gravitational constant. The mass loss rate can be parametrised as  \cite{MacGibbon:1991tj}
\begin{equation}
\frac{dm_\text{BH}(t)}{dt}=-\frac{\mathcal{G}\,g_\star\left(T_\text{BH}\right)}{30720\,\pi}\,\frac{M_\text{pl}^4}{m_\text{BH}^{\rm in}(t)^2}\,,
\label{eq:pbh-dmdt}
\end{equation}
where $\mathcal{G}\sim 4$ is the grey-body factor. Here we ignore the temperature dependence of $g_\star$ during PBH evolution, valid in the pre-sphaleron era. On integrating Eq.~\eqref{eq:pbh-dmdt} we end up with the PBH mass evolution equation as
\begin{equation}
m_\text{BH}(t)=m_\text{BH}(T_\text{in})\Bigl(1-\frac{t-t_\text{in}}{\tau}\Bigr)^{1/3}\,,
\end{equation}
with 
\begin{equation}
\tau = \frac{10240\,\pi\,m_\text{BH}^{\rm in 3}}{\mathcal{G}\,g_\star(T_\text{BH})\,M_\text{pl}^4}\,,
\end{equation}
as the PBH lifetime. Here onward we will use $m_\text{in}(T_\text{in})$ simply as $m_\text{in}$. The evaporation temperature can then be computed taking into account $H(T_\text{evap})\sim\frac{1}{\tau^2}\sim\rho_\text{rad}(T_\text{evap})$ as
\begin{equation}
T_\text{ev}\equiv\Bigl(\frac{45\,M_\text{pl}^2}{16\,\pi^3\,g_\star\left(T_\text{evap}\right)\,\tau^2}\Bigr)^{1/4}\,.
\label{eq:pbh-Tev}
\end{equation}
However, if the PBH component dominates at some point the total energy density of the universe, the SM temperature just after the complete evaporation of PBHs is: $\overline{T}_\text{evap}=2/\sqrt{3}\,T_\text{evap}$~\cite{Bernal:2020bjf}. 

The initial PBH abundance is characterized by the dimensionless parameter $\beta$ that is defined as
\begin{equation}
\beta\equiv\frac{\rho_\text{BH}\left(T_\text{in}\right)}{\rho_\text{rad}\left(T_\text{in}\right)}\,,
\end{equation}
that corresponds to the ratio of the initial PBH energy density to the SM energy density at the time of formation. Note that, $\beta$ steadily grows until PBH evaporation since the PBH energy density scales like non-relativistic matter $\sim a^{-3}$, while the radiation energy density scales as $\sim a^{-4}$. Therefore, an initially radiation-dominated universe will eventually become matter-dominated if the PBHs are still around. The condition of PBH evanescence during radiation domination can be expressed as~\cite{Masina:2020xhk}
\begin{equation}
\beta<\beta_\text{crit}\equiv \gamma^{-1/2}\,\sqrt{\frac{\mathcal{G}\,g_\star(T_\text{BH})}{10640\,\pi}}\,\frac{M_\text{pl}}{m_\text{in}}\,,
\label{eq:pbh-ev-rad}
\end{equation}
where $\beta_c \equiv \beta_{\rm crit}$ is the critical PBH abundance that leads to early matter-dominated era. Note that for simplicity, we consider a monochromatic mass function of PBHs implying all PBHs to have identical masses. Additionally, the PBHs are assumed to be of Schwarzschild type without any spin and charge. The gravitational waves (GW) induced by large-scale density perturbations laid by PBHs could lead to a backreaction problem~\cite{Papanikolaou:2020qtd, Bernal:2020bjf}, that can be avoided if the energy contained in GWs never overtakes the one of the background universe or in other words if\footnote{More recent analysis~\cite{Domenech:2020ssp} shows $8\times 10^{-4}\gtrsim\beta\gtrsim 6\times 10^{-6}$ for PBH of mass $\sim$ 1 g, while $5\times 10^{-10}\gtrsim\beta\gtrsim 10^{-14}$ for PBH of mass $\sim 5\times 10^8$ g.}
\begin{equation}
\beta < 10^{-4}\,\Bigl(\frac{10^9\text{g}}{m_\text{in}}\Bigr)^{1/4}\,.
\end{equation}

Since PBH evaporation produces all particles, including radiation that can disturb the successful predictions of BBN, hence we require $T_\text{evap}>T_\text{BBN}\simeq 4$ MeV. This can be translated into an upper bound on the PBH mass. On the other hand, a lower bound on PBH mass can be obtained from the CMB bound on the scale of inflation \cite{Planck:2018jri} : $\mathcal{H}_I\equiv \mathcal{H}(T_\text{in})\leq 2.5\times 10^{-5}\,M_\text{pl}$, where $\mathcal{H}(T_\text{in})=\frac{1}{2\,t_\text{in}}$ with $t(T_\text{in})=\frac{m_\text{in}}{M_\text{pl}^2\,\gamma}$ (as obtained from Eq.~\eqref{eq:pbh-mass}). Using these BBN and CMB bounds together, we have a window for allowed initial mass for PBH that reads $0.1\,\text{g}\lesssim m_\text{in}\lesssim 3.4\times 10^8\,\text{g}$. The range of PBH masses between these bounds is at present generically unconstrained~\cite{Carr:2020gox}. While PBH can evaporate by Hawking radiation, it can be stable on cosmological scales if sufficiently heavy, potentially giving rise to some or all of DM ~\cite{Carr:2020xqk}. In order not to spoil the structure formation, a DM candidate which is part of the thermal bath or produced from the thermal bath should have mass above a few keV in order to give required free-streaming of DM as constrained from Lyman-$\alpha$ flux-power spectra~\cite{Irsic:2017ixq, Ballesteros:2020adh, DEramo:2020gpr}. Such light DM of keV scale leads to a warm dark matter (WDM) scenario having free-streaming length within that of cold and hot DM. If such light DM is also produced from PBH evaporation, it leads to a potential hot component in total DM abundance, tightly constrained by observations related to the CMB and baryon acoustic oscillation (BAO) leading to an upper bound on the fraction of this hot component with respect to the total DM, depending on the value of DM mass \cite{Diamanti:2017xfo}\footnote{A conservative $10~\%$ upper bound on such hot dark matter (HDM) component \cite{Bernal:2020bjf} can lead to similar constraints on DM mass along with PBH initial fraction.
}. The lower bound on DM mass from Lyman-$\alpha$ can be found in Appendix~\ref{sec:ly-alpha}.
\begin{figure}[htb!]
$$
\includegraphics[scale=0.34]{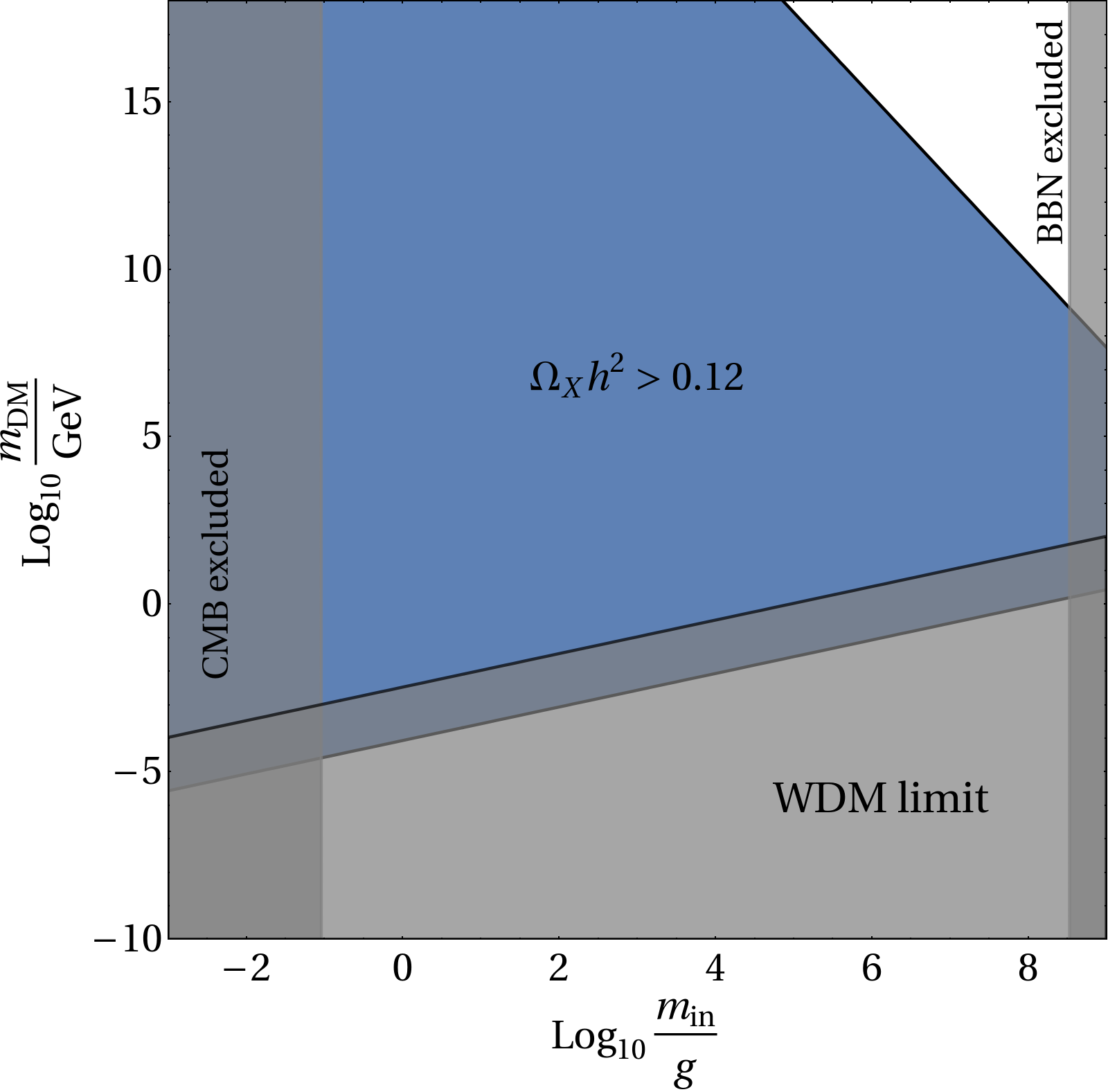}~~~~\includegraphics[scale=0.34]{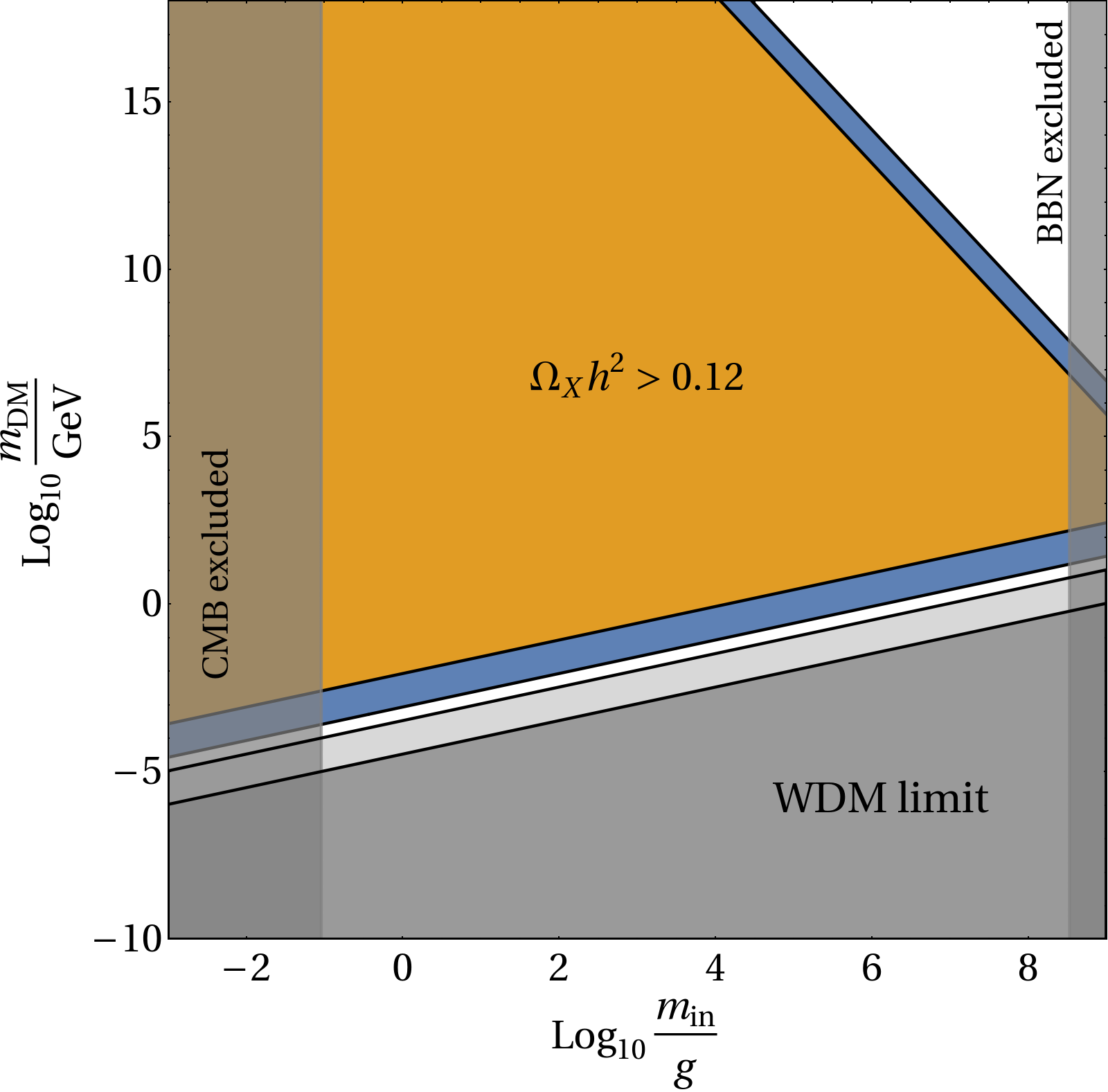}
$$
\caption{The DM is over-abundant in the blue shaded region, for $\zeta=\xi=1$ (left) and in the blue (orange) shaded $\zeta=\xi= 10 (100) $ (right) considering the DM to be scalar. The grey shaded regions are discarded due to limits on PBH mass from CMB (lower limit), BBN (upper limit) and Lyman-$\alpha$ (see text). Observed DM relic is achievable only in the white region in the top right corner (left), whereas non-zero values of $\xi$ opens up a window in the bottom (right). }
\label{fig:overabund}
\end{figure}
We consider DM of spin $s$ with {\it only} gravitational interaction such that it can be produced solely from PBH evaporation. Along with the DM, the PBH also emits right handed neutrinos $N_i$ with $i\in 1,2,3$. The number of any particle $X $ radiated during the evaporation of a single PBH
\begin{equation}
    \mathcal{N}_X = \frac{g_{X,H}}{g_{\star,H}(T_\text{BH})}
    \begin{cases}
       \frac{4\,\pi}{3}\,\Bigl(\frac{m_\text{in}}{M_\text{pl}}\Bigr)^2 &\text{for } m_X < T_\text{BH}^\text{in}\,,\\[8pt]
        \frac{1}{48\,\pi}\,\Bigl(\frac{M_\text{pl}}{m}\Bigr)^2 &\text{for } m_X > T_\text{BH}^\text{in}\,,
    \end{cases}\label{eq:pbh-num}\,,
\end{equation}
where
\begin{equation}
g_{\star,H}(T_\text{BH})\equiv\sum_i\omega_i\,g_{i,H}\,; g_{i,H}=
    \begin{cases}
        1.82
        &\text{for }s=0\,,\\
        1.0
        &\text{for }s=1/2\,,\\
        0.41
        &\text{for }s=1\,,\\
        0.05
        &\text{for }s=2\,,\\
    \end{cases}
\end{equation}
with $\omega_i=2\,s_i+1$ for massive particles of spin $s_i$, $\omega_i=2$ for massless species with $s_i>0$ and $\omega_i=1$ for $s_i=0$. At temperatures $T_\text{BH}\gg T_\text{EW}\simeq 160$ GeV, PBH evaporation emits the full SM particle spectrum according to their $g_{\star,H}$ weights, while at temperatures below the MeV scale, only photons and neutrinos are emitted. For $T_\text{BH}\gg 100$ GeV (corresponding to $m_\text{BH}\ll 10^{11}$ g), the particle content of the SM corresponds to $g_{\star,H}\simeq 108$. The DM yield produced by evaporation is directly related to the BH abundance at evaporation as
\begin{equation}
Y_\text{DM}(T_0) = \frac{n_\text{DM}}{s}\Big|_{T_0}=\,\mathcal{N}_X\,\frac{n_\text{BH}}{s}\Big|_{T_\text{ev}}\,.
\end{equation}
Now, PBH abundance at evaporation for PBH domination can be obtained using the first Friedmann equation as
\begin{equation}
n_\text{BH} (T_\text{ev}) = \frac{1}{6\,\pi}\,\frac{M_\text{pl}^2}{m_\text{in}^2\,\tau^2}\equiv\frac{1}{6\,\pi}\,\left(\frac{\mathcal{G}\,g_{\star,H}}{10640\,\pi}\right)^2\,\frac{M_\text{pl}^{10}}{m_\text{in}^7}\,.  
\end{equation}
Then, the DM relic abundance $\Omega_\text{DM}\,h^2 = \frac{m_\text{DM}\,s_0}{\rho_c}\,Y_\text{DM}\left(T_0\right)\,,$ in the present epoch reads 
\begin{equation}
    \Omega_\text{DM}\,h^2 =\mathbb{C}(T_\text{ev})
    \begin{cases}
      \frac{1}{\pi^2}\,\sqrt{\frac{M_\text{pl}}{m_\text{in}}}\,m_\text{DM} &\text{for } \mdm < \Tin\,,\\[8pt]
       \frac{1}{64\,\pi^4}\left(\frac{M_\text{pl}}{m_\text{in}}\right)^{5/2}\,\frac{M_\text{pl}^2}{m_\text{DM}} &\text{for } \mdm > \Tin\,,
    \end{cases}\label{eq:rel-dm}\,
\end{equation}
with $\mathbb{C}(T_\text{ev})=\frac{s_0}{\rho_c}\,\frac{1}{\zeta}\,\frac{g_{X,H}}{g_{\star,H}}\,\frac{5}{g_{\star s}(T_\text{ev})}\,\,\left(\frac{\pi^3\,g_{\star}(T_\text{ev})}{5}\right)^{3/4}\,\sqrt{\frac{\mathcal{G}\,g_{\star,H}}{10640\,\pi}}$. Here $\zeta$ parametrizes a possible entropy production after PBH evaporation until now, i.e., $\zeta\,\left(sa^3\right)_\text{evap}=\left(sa^3\right)_0$. Note that, for the heavy DM case relic abundance has an inverse dependence on the DM mass, implying heavy DM leads to under abundance. The over abundant region for DM with spin $s=0$ produced from PBH evaporation is shown in the left panel of Fig.~\ref{fig:overabund} for $\zeta=\xi=1$, i.e., considering no entropy injection at any given epoch, {{where $\xi$ is defined in Appendix~\ref{sec:ly-alpha} and for our scenario $\xi = \zeta$ as there is no entropy dilution after matter-radiation equality epoch.}} Here we note that in the majority of the parameter space the DM is over abundant irrespective of their spins and only $\mdm\gtrsim 10^{10}$ GeV can lead to right abundance for $\Min\gtrsim 10^6$ g~\cite{Fujita:2014hha, Samanta:2021mdm}. Here we would like to mention that for DM with different spins this over abundant region only slightly changes. Particularly, the low DM mass region is in conflict with the WDM limit. An obvious way to overcome this tension is to consider that the DM produced from PBH evaporation does not constitute the whole DM abundance. But since we are considering PBH is the {\it only} source of all of the DM, therefore this tension can be alleviated by considering entropy non-conservation~\cite{Fujita:2014hha, Masina:2020xhk}. This is shown in the right panel of Fig.~\ref{fig:overabund}, where $\zeta=\xi=10$ (in blue) and $\zeta=\xi=100$ (in yellow) is assumed\footnote{Another way is to significantly increase $g_{\star,H}$ as mentioned in~\cite{Masina:2020xhk}.}. Here we see the low DM mass window can be resurrected simply because $\Omega_\text{DM}\propto\,\zeta^{-1}\,\mdm$\footnote{For simplicity we consider $\zeta=\xi$, which is, of course not a mandate.}. 
\begin{figure}[htb!]
$$
\includegraphics[scale=0.35]{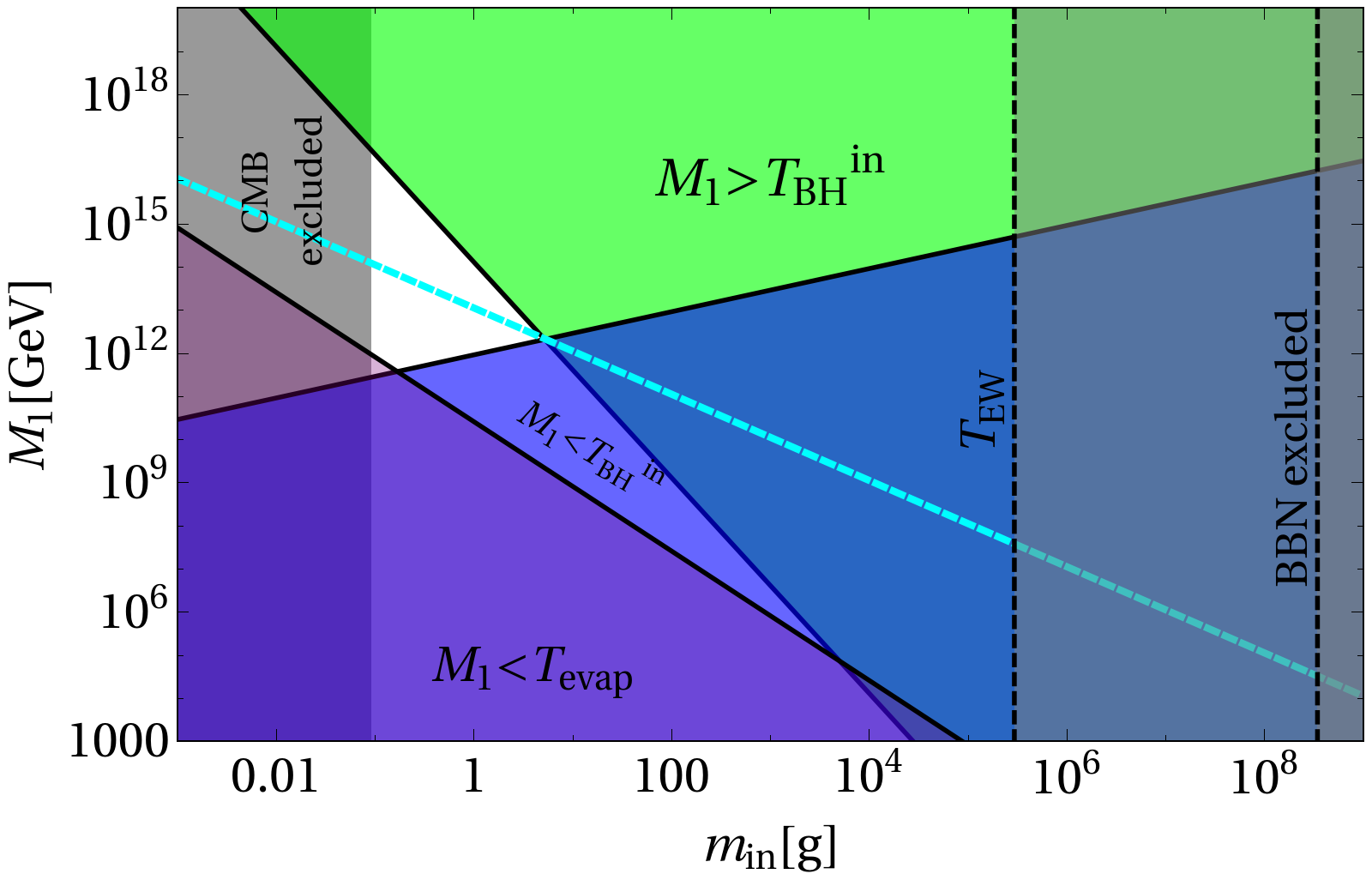}~~~~\includegraphics[scale=0.35]{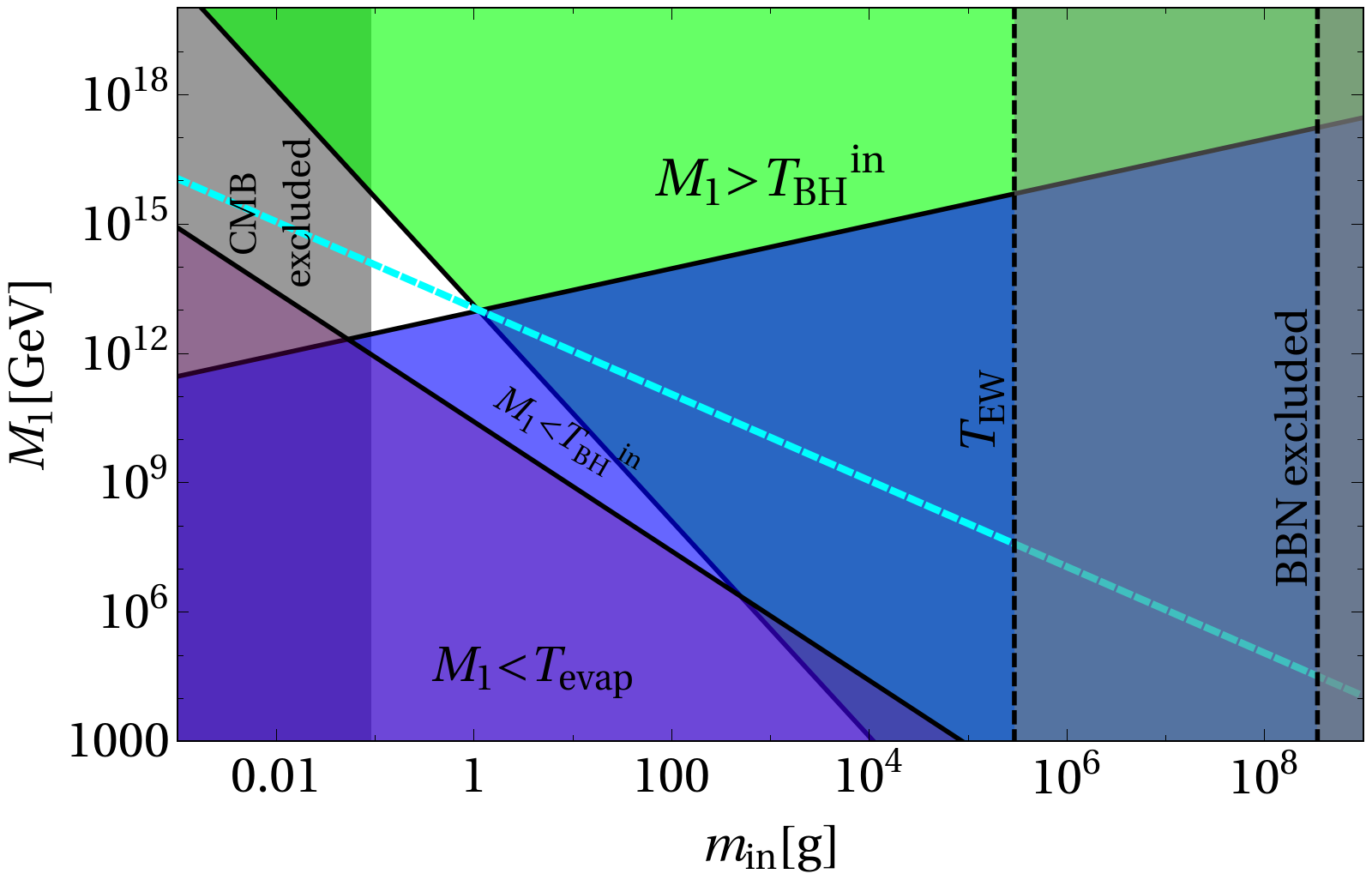}
$$
\caption{Constraints on RHN mass from the requirement of obtaining $Y_B^\text{obs}$ considering $\zeta=1$ (left) and $\zeta=10$ (right). All the coloured regions are discarded from the bounds derived in Eq.~\eqref{eq:m1bound}. The vertical black dashed line corresponds to (from left to right) the bound from the scale of inflation (CMB), sphaleron transition and BBN. The cyan dashed diagonal straight line corresponds to $\Tin=\Min$. The white triangular region in the middle is the region that is allowed (see text).}
\label{fig:m1-bound}
\end{figure}
The RHNs emitted during PBH evaporation can undergo CP-violating decays, generating lepton asymmetry. This lepton asymmetry is then further converted into the observed baryon asymmetry of the Universe via sphaleron transition~\cite{Fujita:2014hha}. It is possible to analytically derive the mass range of RHNs (and PBH) emitted from PBH evaporation that can provide the observed baryon asymmetry. In the Type-I seesaw mechanism, the quantity $\epsilon$ has an upper bound~\cite{Davidson:2002qv,Samanta:2020gdw}
\begin{equation}\label{eq:cp-asym}
\epsilon\lesssim \frac{3}{16\,\pi}\,\frac{M_1\,m_{\nu,\text{max}}}{v^2}\,,
\end{equation}
where $v=246$ GeV is the SM Higgs VEV and $m_{\nu,\text{max}}$ is the mass of the heaviest light neutrino. On the other hand, the final asymmetry produced from PBH evaporation is given by $Y_B^\text{obs}=n_B/s\Big|_{T_0}=\frac{1}{\zeta}\,\mathcal{N}_X\,\epsilon\,a_\text{sph}\,Y_B\Big|_{\Tev}\simeq 8.7\times 10^{-11}$~\cite{Planck:2018jri}, where $a_\text{sph}\simeq 1/3$ and $T_0$ is the present temperature of the Universe. These together constrain the mass of the RHN  produced from PBH evaporation both from above and from below~\cite{Fujita:2014hha}
\begin{equation}
    M_1 
    \begin{cases}
        > \frac{4\,g_{\star,H}(\Tin)}{g_X\,a_\text{sph}}\,\zeta\,\frac{Y_B^0}{Y_B^\text{evap}}\,\frac{v^2\,M_\text{pl}^2}{m_\nu\,\Min^2} &\text{for } M_1 < \Tin\,;\\[8pt]
        < \frac{g_X\,a_\text{sph}}{256\,\pi^2\,g_{\star,H}}\,\frac{1}{\zeta}\,\frac{Y_B^\text{evap}}{Y_B^0}\,\frac{M_\text{pl}^2\,m_\nu}{v^2} &\text{for } M_1 > \Tin\,,
    \end{cases}\label{eq:m1bound}
\end{equation}
To ensure non-thermal production of baryon asymmetry it is also necessary to consider $M_1 > T_\text{evap}$~\cite{Fujita:2014hha} that leads to
\begin{equation}
M_1 \gtrsim 3\times 10^{-3}\,\left[\mathcal{G}^2\,g_\star(\Tev)\left(\frac{M_\text{pl}^5}{\Min^3}\right)^2\right]^{1/4}\,,
\label{eq:m1bound2}
\end{equation}
otherwise for $M_1<T_\text{evap}$, the RHNs produced from PBH evaporation are in thermal bath and then washout processes are in effect. Finally, in order for lepton asymmetry to be sufficiently generated from RHNs produced from PBH evaporation, one requires evaporation to be over before sphaleron transition $\Tev\gtrsim T_\text{EW}$, which translates into $\Min\lesssim 3\times 10^5\,\text{g}$. In Fig.~\ref{fig:m1-bound} we summarize the bounds on RHN mass, required to produce the observed baryon asymmetry. The tiny triangular white part is the only window where $Y_B=Y_B^\text{obs}$. This region typically corresponds to $0.1\lesssim \Min\lesssim 20$ g and $10^{12}\lesssim M_1\lesssim 10^{17}$ GeV when entropy is assumed to be conserved, i.e., $\zeta=1$. Note that, this region shrinks for larger $\xi=10$, as shown in the right panel. This is however expected since a larger $\zeta$ allows a larger entropy injection (from $\Tev$ to $T_0$), diluting the asymmetry produced. Thus, while a larger $\zeta$ can provide  breathing space for lighter DM (cf., Fig.~\ref{fig:overabund}), but in turn tightens the allowed parameter space for observed baryon asymmetry. It is therefore clear, satisfying both of them simultaneously needs a careful choice of $\zeta$, i.e., the entropy injection, such that not only right DM abundance for lighter DM is obtained, but baryon asymmetry should also not get too much diluted. It is interesting to note that the heavy DM mass $(\gtrsim 10^{10}~\rm GeV)$ region although remains viable even in the absence of entropy injection, it is not possible to satisfy the observed asymmetry in those regions as they typically correspond to massive PBH $(\gtrsim 10^4~\rm g$), leading to under production of asymmetry (cf. Fig.~\ref{fig:m1-bound}).

\section{Numerical Analysis}
\label{sec:numerical}
To this end we have analytically established that in order to open up the low mass DM window (which is otherwise over abundant) it is necessary to have a substantial entropy injection $(\gtrsim\mathcal{O}(10))$. However, the same entropy injection reduces the viable parameter space for observed baryon asymmetry. Thus, it is rather difficult to satisfy both DM abundance and correct asymmetry together, and one has to stick to the region of comparatively lighter PBH mass to achieve both. In this section we will investigate the viability of the analytical results by solving a set of coupled Boltzmann equations (BEQ) numerically.

\begin{figure}[htb!]
$$
\includegraphics[scale=0.27]{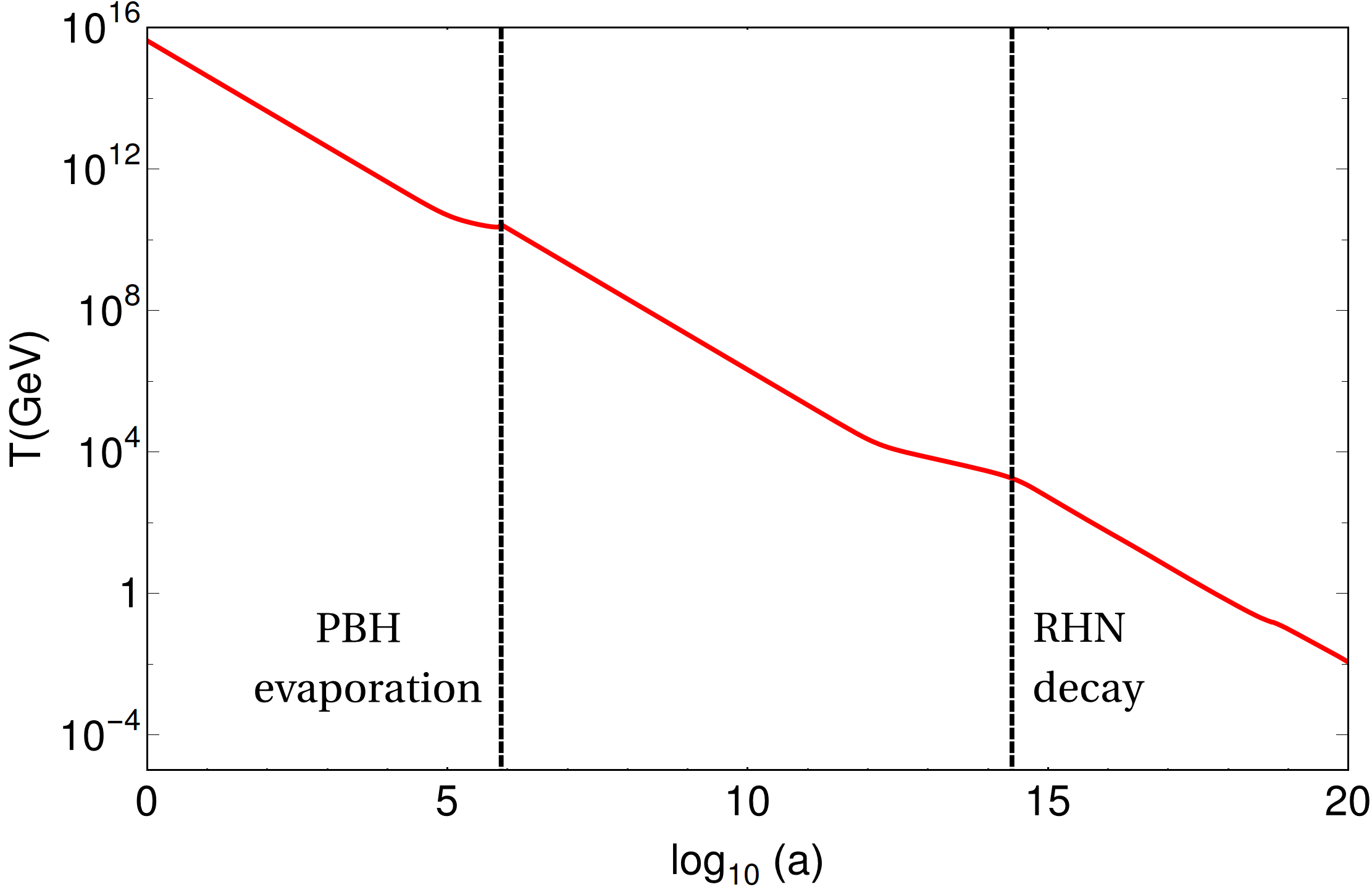}~~~~
\includegraphics[scale=0.27]{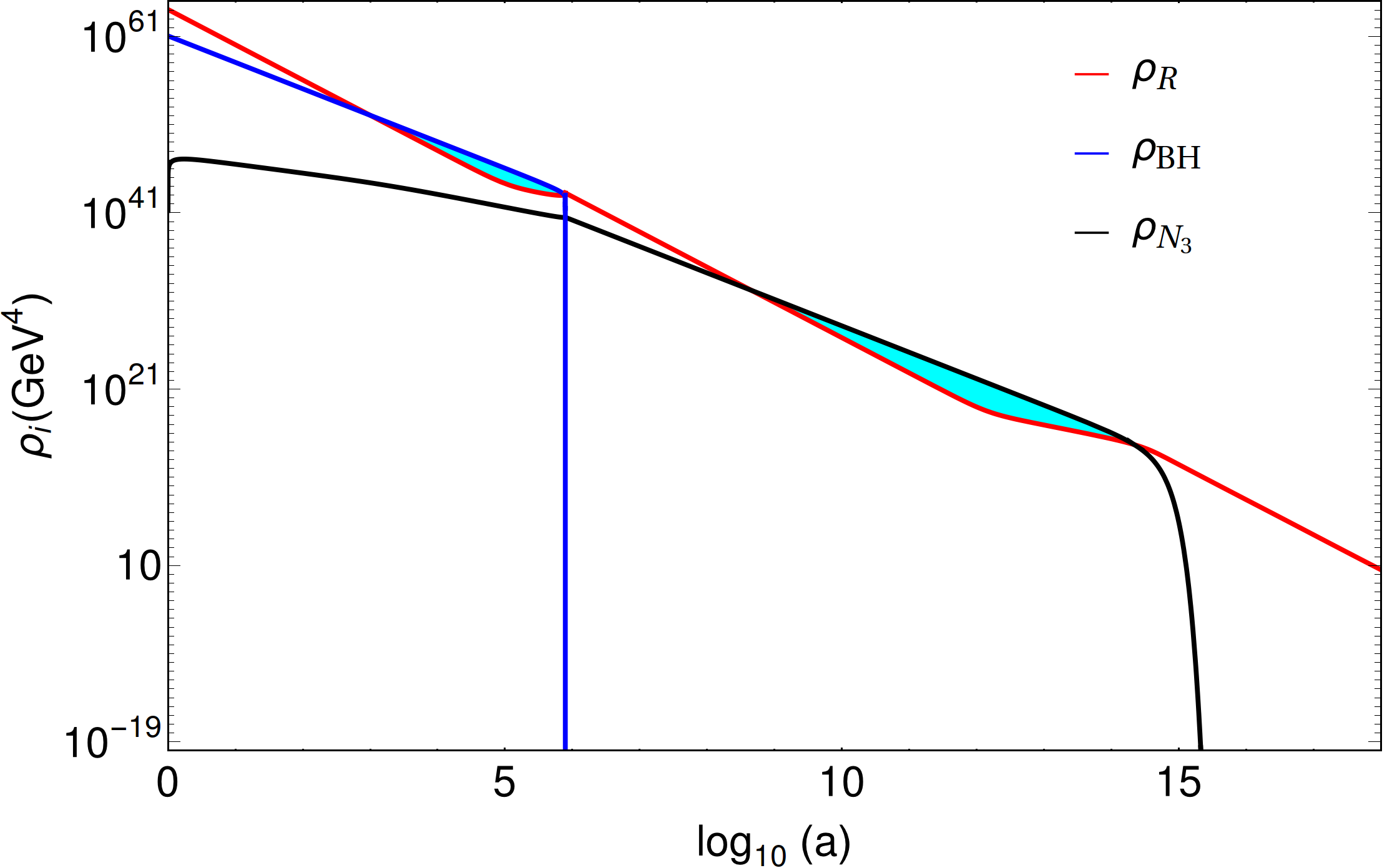}
$$
\caption{ Evolution of temperature of the thermal bath $T$ as a function of the scale factor $a$ (left) and  energy densities of radiation, PBH and $N_3$ as a function of the scale factor (right). We take $m_{in}=1$ g, $M_1 = 10^{13}$ GeV, $M_3 = 10^{12}$ GeV, $m_\text{DM} = 1$ GeV, with $N_3$ decay width adjusted to be $\Gamma_3 = 1.3\times 10^{-11}$ GeV (see text). }
\label{fig:entdil1}
\end{figure}

In order to compute the final lepton (baryon) and DM yield, we numerically solve the set of coupled BEQs in Appendix~\ref{sec:cBEQ}. Since we are interested in PBHs with mass $\lesssim\mathcal{O}(1)$ g (where leptogenesis from PBH dominates), with typical evaporation temperature $T_\text{evap}\lesssim\mathcal{O}(10^{11})$ GeV, we do not include the $\Delta L=2$ washout processes in the BEQ as such processes go out of equilibrium at temperatures $T\lesssim 6\times 10^{12}$ GeV as shown in~\cite{Bernal:2022pue} and thus have no influence on final asymmetry\footnote{ In the standard radiation dominated early Universe, if the lightest RHN mass
exceeds $\sim 10^{15}$ GeV and heaviest active neutrino mass is greater than $\sim 0.1$ eV, $\Delta L=2$ washout processes erase the lepton asymmetry.}. We consider $\beta= 10^{-3}$, such that PBH dominate the energy density at some epoch and particle production takes place during PBH domination. The temperature of the thermal bath as a function of the scale factor is shown in the left panel of Fig.~\ref{fig:entdil1}, where we can clearly see the effect of entropy dilution in two different epochs. The first one takes place at $a\sim 10^6$ when the PBH evaporation is completed, while the second one at a later epoch corresponding to $a\sim 10^{15}$ when the decay of $N_3$ is completed. Note that, prior to $N_3$ domination, the universe is again dominated by radiation energy density for a brief period of time.  This is clearly visible from the right panel plot where we have shown the energy density of radiation (red), PBH (blue) and $N_3$ (black) as functions of the scale factor $a$. The cyan-shaded regions indicate the two different epochs of early matter domination. Here one can see that the PBH energy density falls sharply at $a\sim 10^6$ denoting the end of PBH domination. Afterward the Universe goes through radiation domination, that is being overtaken by the $N_3$ energy density (second matter dominated era) at $a\sim 10^9$. Finally that ends at $a\sim 10^{15}$ with the complete decay of $N_3$ into radiation. It is important to clarify that in these plots we have fixed the $N_3$ mass and adjusted its decay width accordingly such that it never achieves equilibrium with the SM bath (hence long-lived) by tuning the lightest active neutrino mass that we consider to be a free parameter. 

\begin{figure}[htb!]
$$
\includegraphics[scale=0.27]{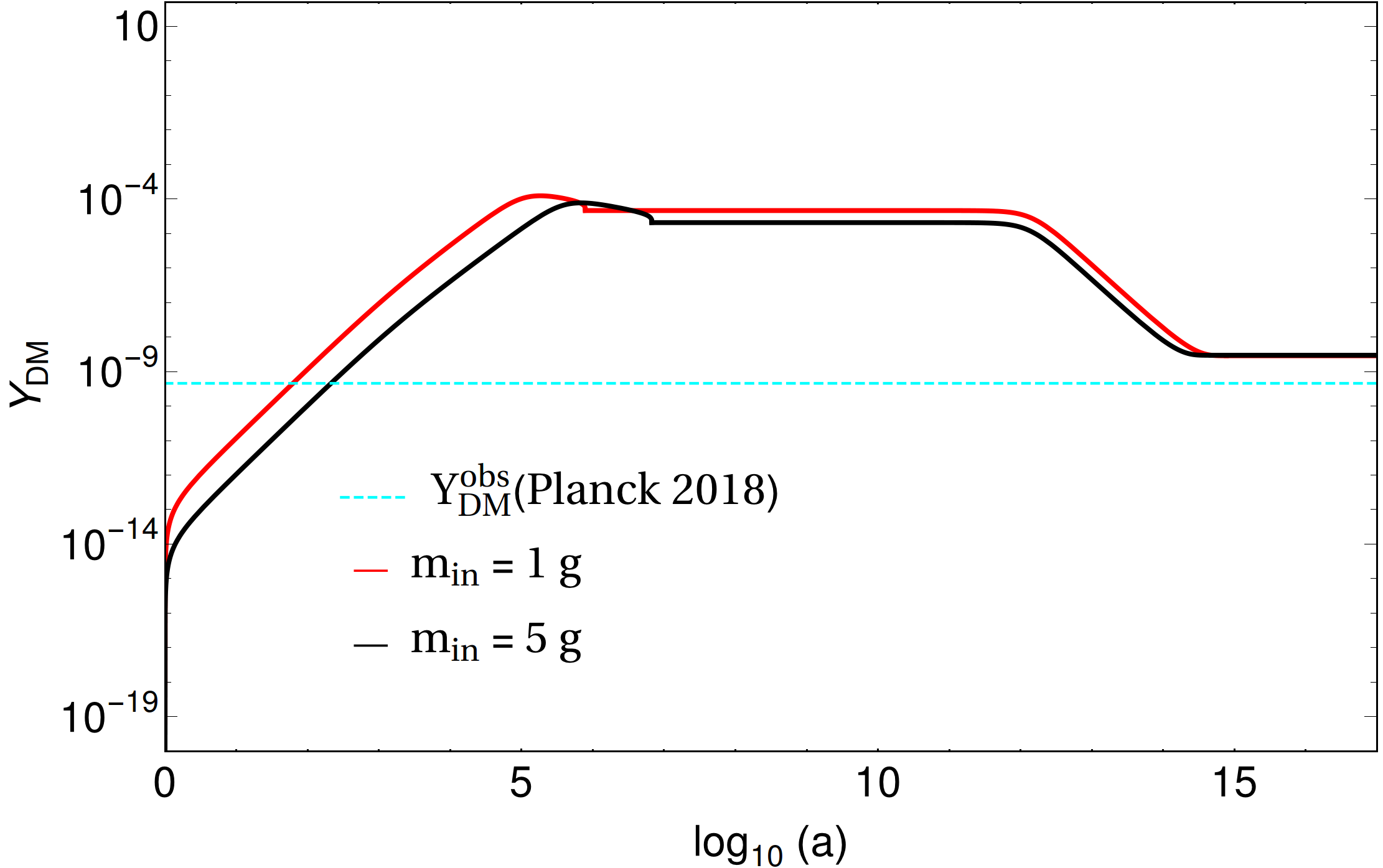}~~~~
\includegraphics[scale=0.27]{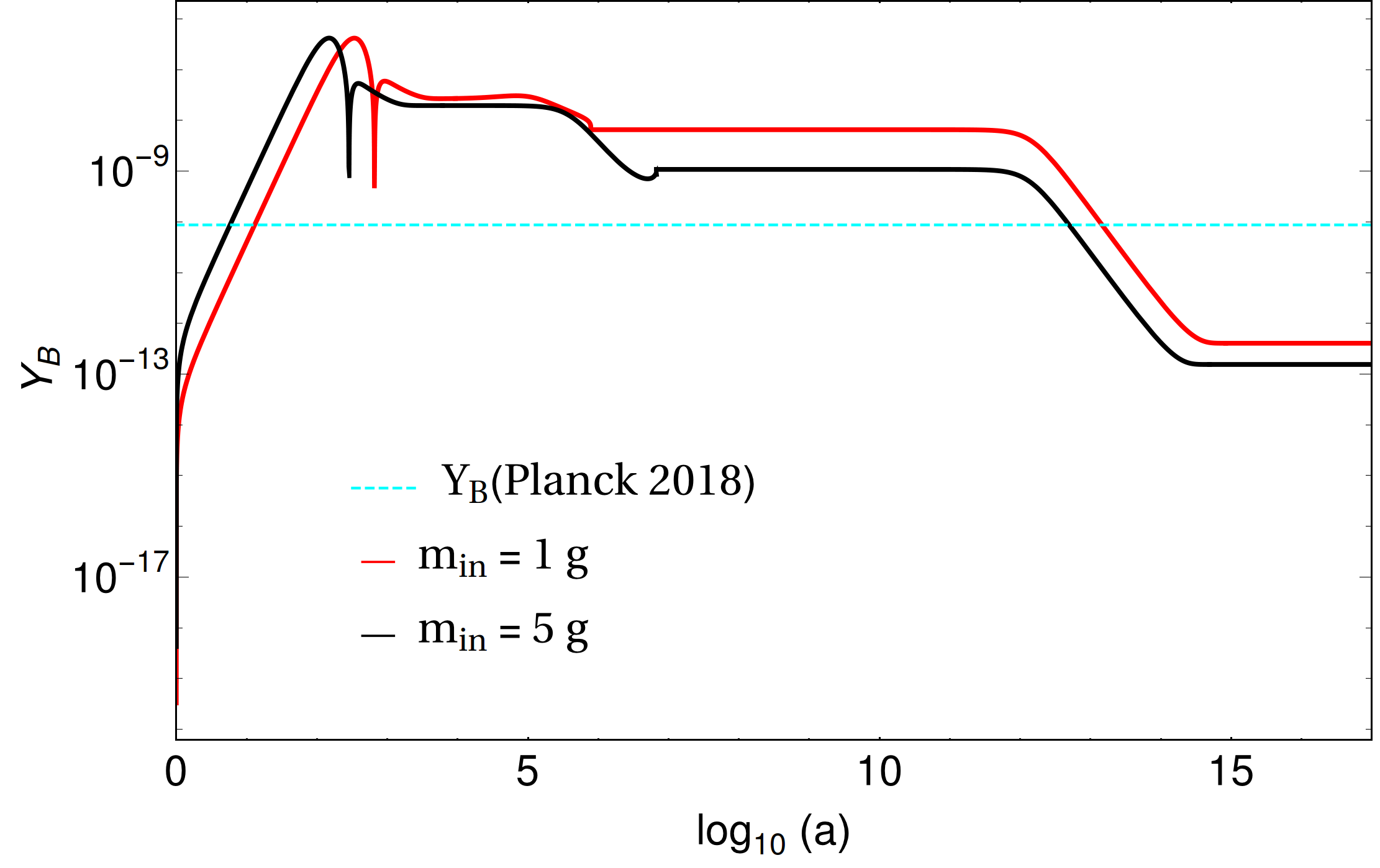}
$$
\caption{Evolution of DM (left) and baryon yield (right panel) for two different values of PBH masses shown in two different colours. We consider $m_\text{DM}=1$ GeV, $M_1 = 10^{13}$ GeV,  $M_3=10^{12} $ GeV, while keeping $N_3$ decay $\Gamma_{3} = 1.3 \times 10^{-11}$ GeV to be fixed.}
\label{fig:entdil2}
\end{figure}

\begin{figure}[htb!]
$$
\includegraphics[scale=0.27]{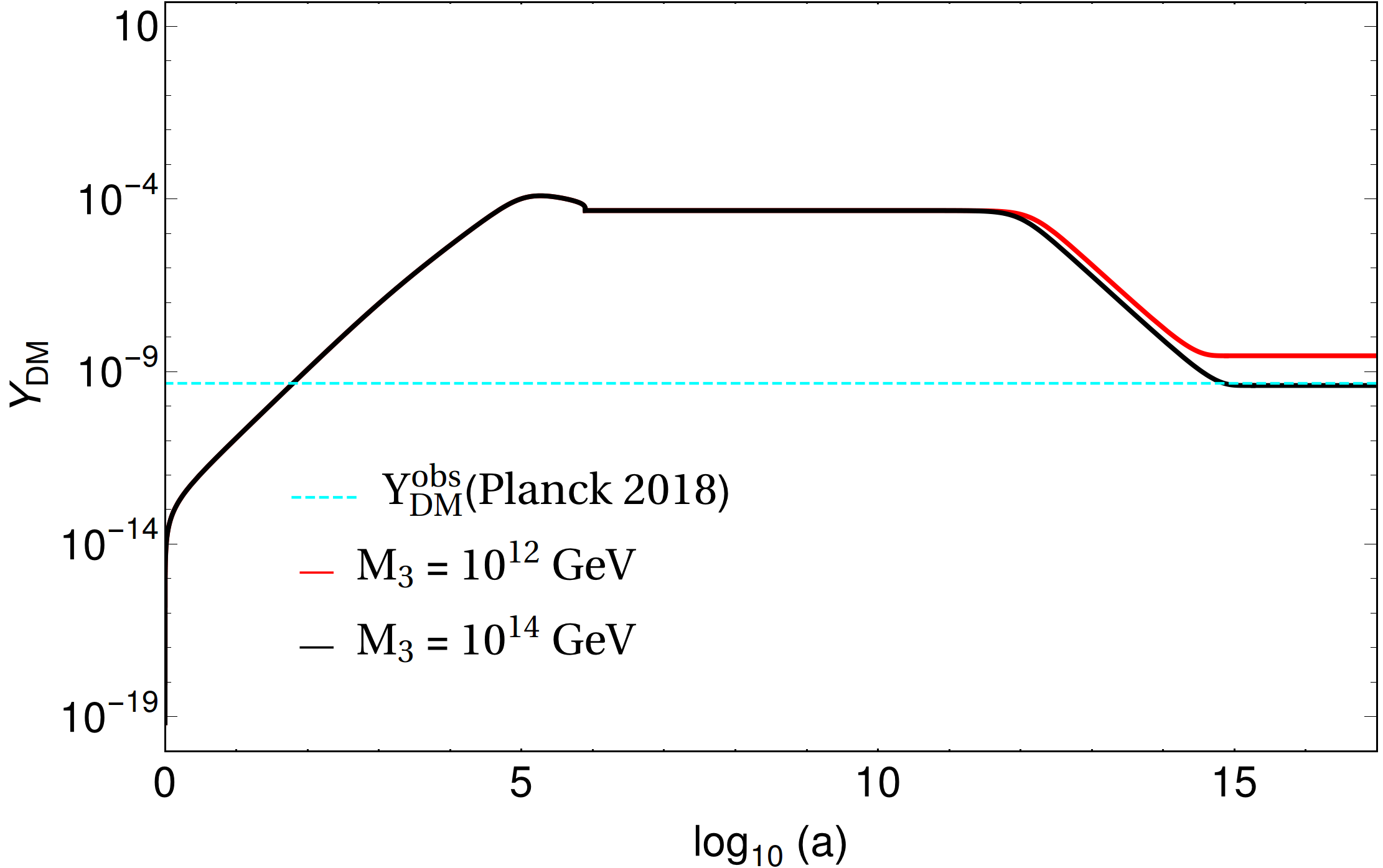}~~~~
\includegraphics[scale=0.27]{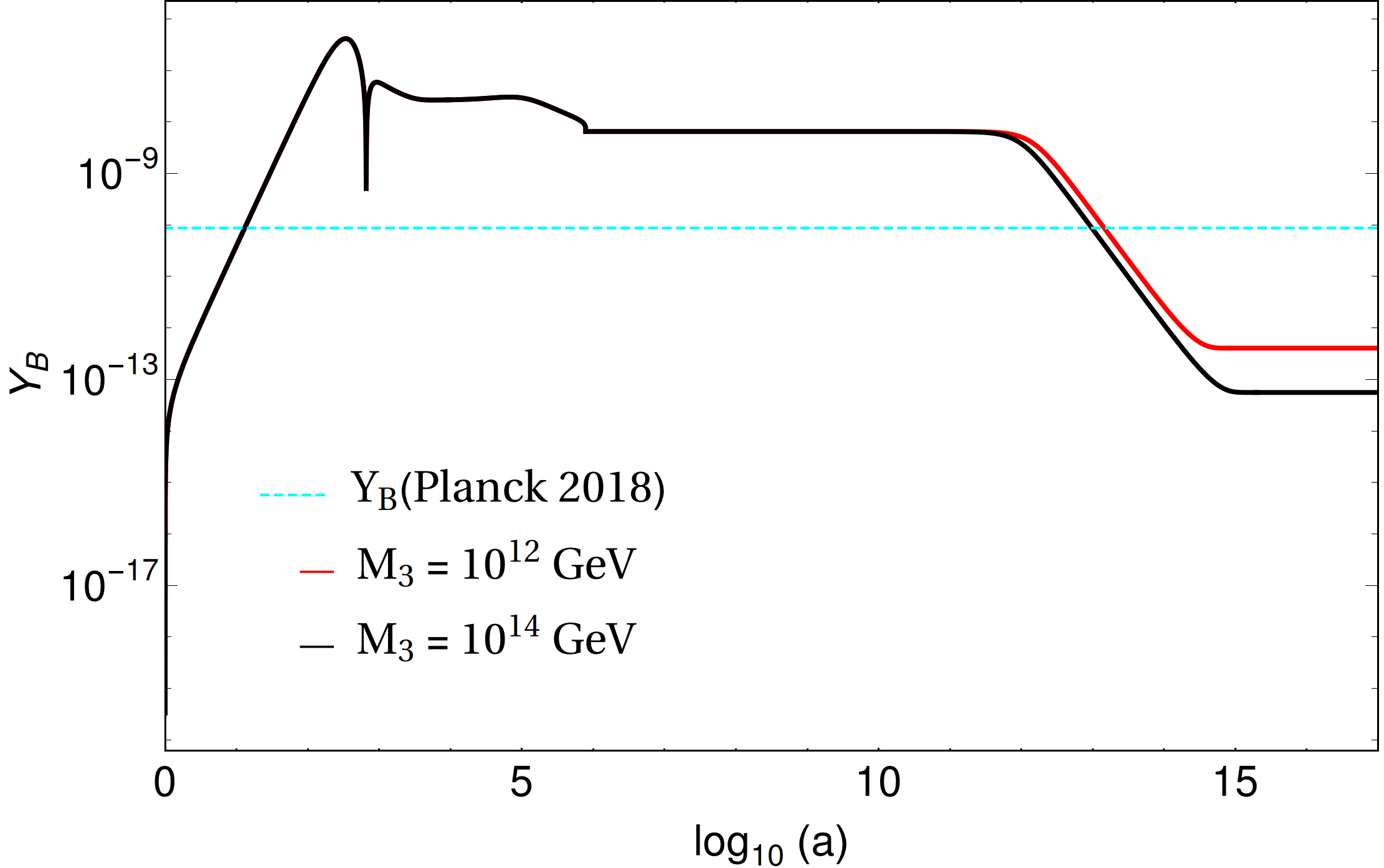}
$$
\caption{Evolution of DM (left) and baryon yield (right panel) for two different $M_3$ values, shown in two different colours. We consider $m_\text{DM}=1$ GeV, $m_\text{in} = 1$ g, and the decay width $\Gamma_{3} = 1.3 \times 10^{-11}$ GeV is kept fixed.}
\label{fig:entdil3}
\end{figure}

\begin{figure}[htb!]
$$
\includegraphics[scale=0.27]{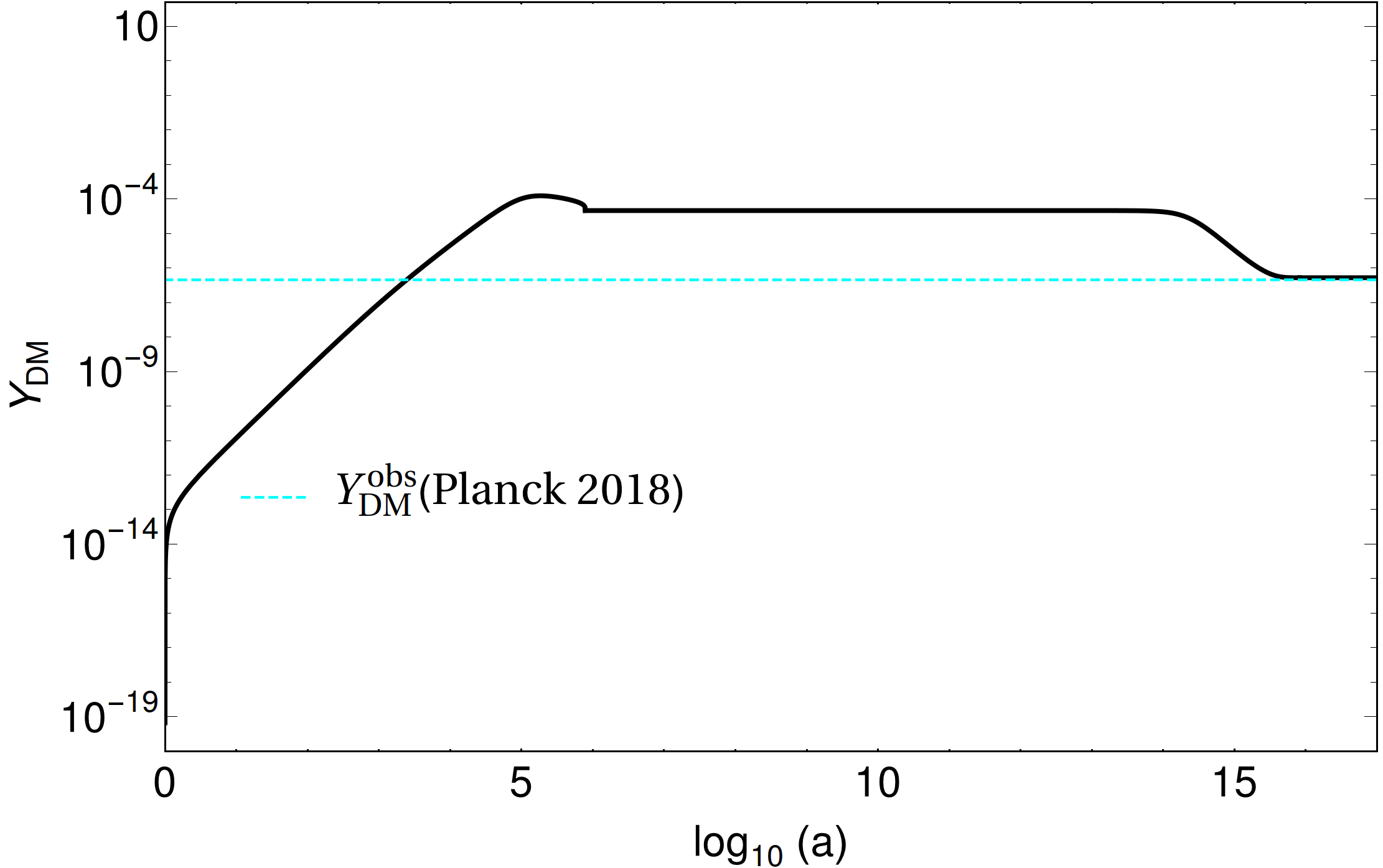}~~~~
\includegraphics[scale=0.27]{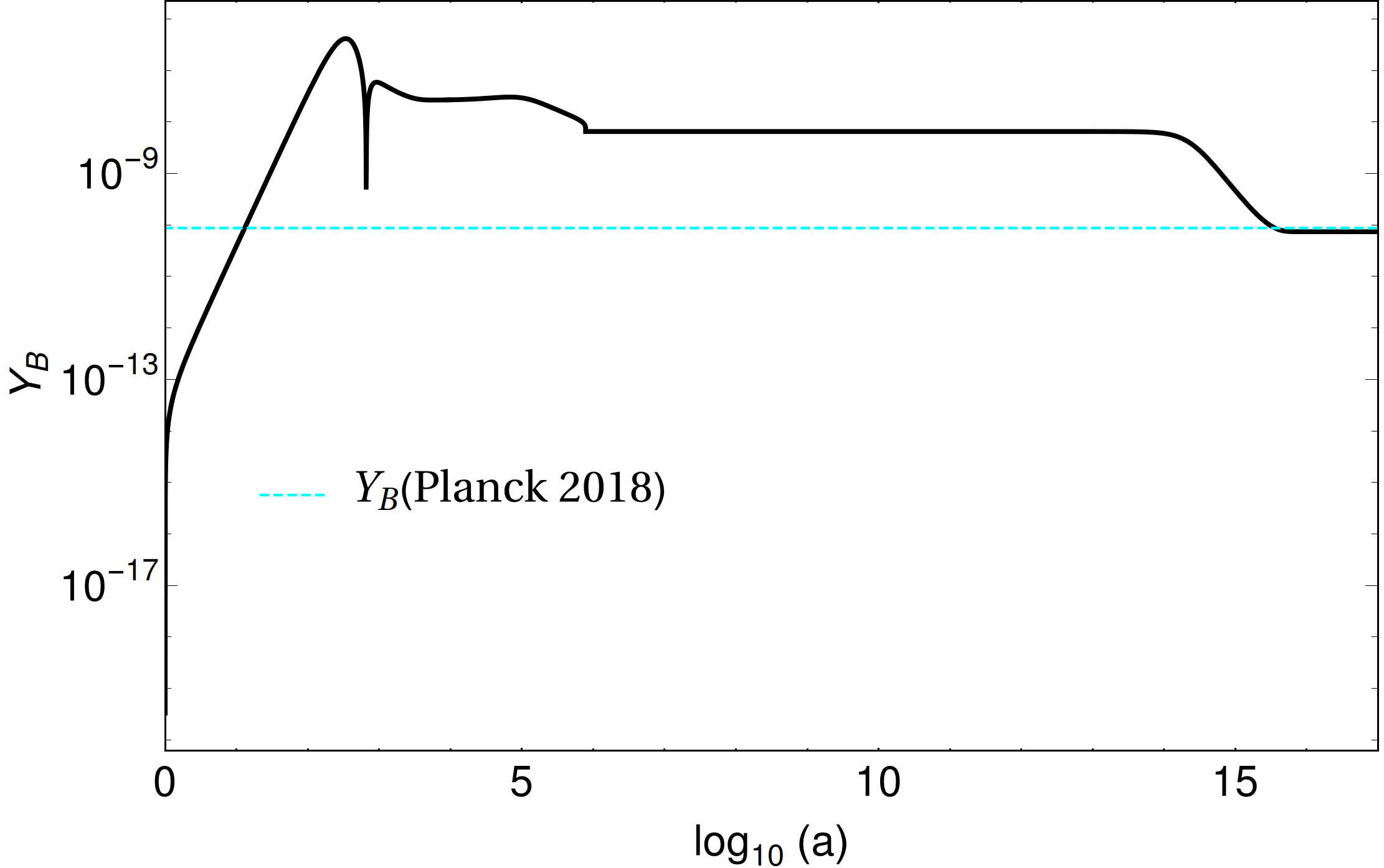}
$$
\caption{Evolution of DM yield (left panel) and baryon asymmetry yield (right panel) for choice of parameters which together satisfy observed DM relic and baryon asymmetry. We choose $m_\text{DM}=1$ MeV, $M_1 = 10^{13}$ GeV, $M_3 = 10^{8}$ GeV, $m_{in} = 1 $g, and  $\Gamma_{3} = 4 \times 10^{-15}$ GeV.}
\label{fig:entdil4}
\end{figure}
As advocated in the last section, DM overproduced from PBH evaporation gets diluted to right abundance by introducing entropy injection at late epochs. In this case $N_3$ decay is responsible for adequate entropy dilution. The effect of entropy injection on the DM and baryon yield is demonstrated in Fig.~\ref{fig:entdil2}, considering two different PBH masses. In the left panel we see DM is first overproduced at the end of PBH evaporation around $a\sim 10^6$. The DM yield then starts diminishing once $N_3$ starts decaying at later epoch near $a\sim 10^{12}$. Once $N_3$ decay is complete, the DM yield saturates close to the Planck 2018 limit for $m_\text{DM}=1$ GeV. As we already noticed in Fig.~\ref{fig:m1-bound}, entropy injection has a destructive effect on the baryon asymmetry since it dilutes the asymmetry generated from RHN decay as well. This is again established in the right panel, where we see the generated yield for asymmetry becomes under-abundant after the completion of $N_3$ decay for a fixed $M_3$. For a fixed PBH mass the effect of varying $M_3$ is shown in Fig.~\ref{fig:entdil3}, where in the left panel we see that a heavier $M_3$ is capable of producing the right DM abundance for a fixed DM mass since a heavier $N_3$ results in larger entropy dilution. This affects $Y_B$ as well, as one can see from the right panel of Fig.~\ref{fig:entdil3}. Therefore, one has to make a careful choice of PBH and diluter mass in order to satisfy both DM abundance and right baryon asymmetry. In Fig.~\ref{fig:entdil4} we show a specific benchmark point which gives rise to right asymptotic yield for the DM and that of baryon asymmetry simultaneously, in agreement with observations. Here we choose a DM of mass $m_\text{DM}=1$ MeV, while $M_3=10^8$ GeV with a PBH of mass 1 g.
\begin{figure}[htb!]
    \centering
    \includegraphics[scale=0.31]{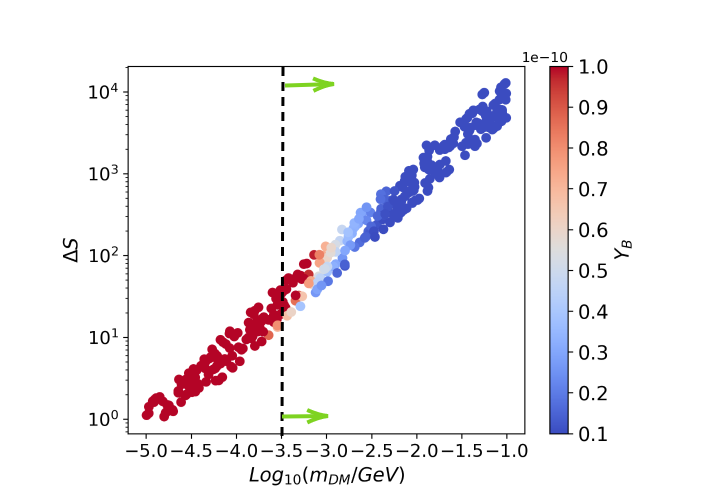} \includegraphics[height=5.61cm,width=6.25cm]{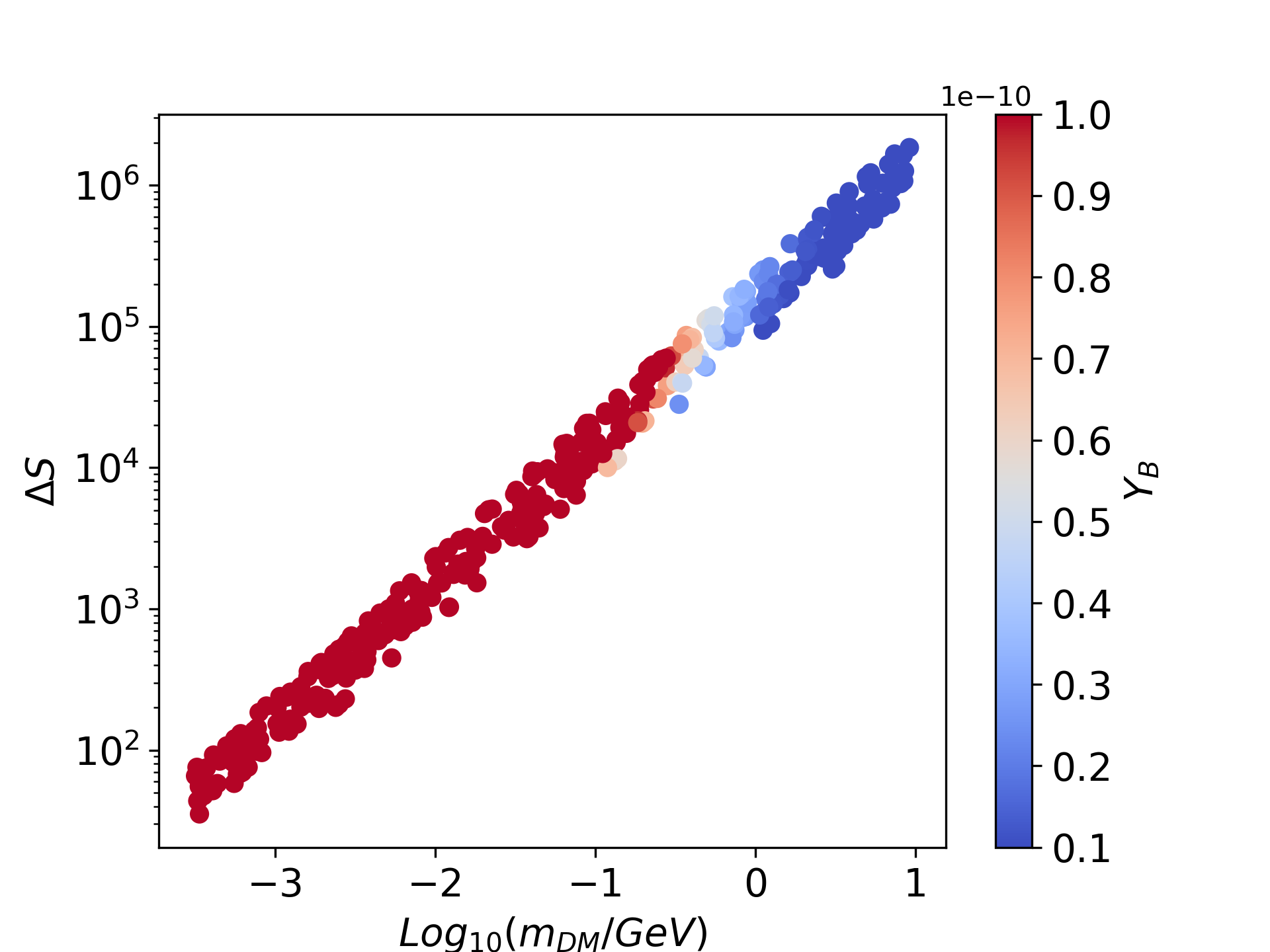}
    \caption{Left: Parameter space allowed by relic density in $\Delta S-m_\text{DM}$ plane, where PBH mass is scanned over the range: \{0.5-5\} g. The colour code is with respect to $Y_B$, considering vanilla high scale leptogenesis. The black vertical dashed line corresponds to the conservative bound from WDM. The green arrows denote the net allowed parameter space. Right: Same as left, but considering resonant leptogenesis.}
    \label{fig:scan}
\end{figure}
The entropy released due to the decay of long-lived heavy $N_3$ can be expressed analytically as~\cite{PhysRevD.31.681, Bezrukov:2009th}
\begin{equation}
\Delta S\simeq \left[1+2.95\,\left(\frac{2\,\pi^2\,g_{\star s}(T)}{45}\right)^\frac{1}{3}\,\left(\frac{Y_3^2\,M_3^2}{M_\text{pl}\,\Gamma_3}\right)^\frac{2}{3}\right]^\frac{3}{4}\,, 
\end{equation}
where $Y_3=n_{N_3}/s$ is the initial yield of $N_3$ before the onset of the second matter dominated era due to $N_3$ and $\Gamma_3$ is its decay width. Note that the amount of entropy injection is inversely proportional to $\Gamma_3$ as expected. Now, $\Gamma_3$ cannot be arbitrarily small as $N_3$ has to decay before BBN. This gives an upper bound on $\Delta\,S$, around $10^{11}$, which translates into an upper bound on the dark matter mass, $m_{\rm DM} \lesssim 10^{6}$ GeV. This is nevertheless a significant improvement as DM masses all the way till $m_{\rm DM} \sim 10^{10}$ GeV, if solely produced from PBH evaporation, remain disallowed in the usual scenario without late entropy injection, either due to overproduction or WDM limits, as shown in the left panel plot of Fig.~\ref{fig:overabund}. It is important to specify here that $\Delta S$ can be related to the lightest neutrino mass $m_\nu^1$,  since the value of the $N_3$ Yukawas are determined by $m_\nu^1$. For instance, we find that for $m_{\rm in}=1$ g, and $M_3 < T_{\rm BH}^{\rm in}$, $1 \lesssim \Delta S \lesssim 10^{6}$ typically corresponds to lightest neutrino mass in the range $10^{-10} \, {\rm eV} \lesssim m_{\nu}^1 \lesssim 10^{-24} \, {\rm eV}$. Such tiny values of the lightest neutrino mass and hence tiny Dirac Yukawa couplings of $N_3$ are expected since $N_3$ has to be sufficiently long-lived. Now, in order to explore the viable parameter space we perform a numerical scan over the DM mass $m_\text{DM}$ in keV-GeV range and on the PBH mass $m_\text{in}:\{0.5-5\}$ g, fixing the leptogenesis scale $M_1$ at $10^{13}$ GeV. In the left panel of Fig.~\ref{fig:scan} we show the relic density allowed parameter space in $\Delta S-m_\text{DM}$ plane. { DM particles with large velocity, i.e., warm DM, are constrained by observations because their large free-streaming length prevents structure formation of the universe. The formation of structures in WDM is suppressed for perturbations of comoving size $\lesssim \lambda_\text{DM}\propto m_\text{WDM}^{-4/3}$~\cite{Drewes:2016upu, Garzilli:2019qki}. The signature of such WDM would thus be the suppression of the matter power spectrum (MPS) at scales below their free-streaming horizon. From cosmological data at large scales (CMB and galaxy surveys) we know that such a suppression should be sought at comoving scales well below a Mpc. The Lyman-$\alpha$ forest has been used for measuring the matter power spectrum at such scale~\cite{Croft:2000hs,GARZILLI2017258}. This lower bound on the mass of a thermal early decoupled WDM can be translated into a lower bound on the present velocity of a generic WDM.} As already established earlier, light DM with mass below $\lesssim 3$ keV~\cite{Viel:2013fqw,Baur:2015jsy} is disallowed from the Lyman-$\alpha$ bound on WDM\footnote{A more conservative bound on WDM mass has been derived in~\cite{Garzilli:2019qki}.}, which are shown by the red points on the left side of black vertical dashed line. However, a substantial amount of entropy injection can improve this situation as already shown in the right panel plot of Fig.~\ref{fig:overabund}. Thus, DM mass $m_\text{DM}\gtrsim \mathcal{O}$(MeV) is allowed for $\Delta S\gtrsim\mathcal{O}(10)$. The parameter space is also capable of explaining the observed baryon asymmetry in vanilla leptogenesis framework for $10\lesssim\Delta S\lesssim 100$, which agrees with earlier observations in~\cite{Fujita:2014hha}. We project the most conservative WDM bound that disallows DM mass $m_\text{DM}\lesssim 0.3$ MeV. Note that, as foretold, with the increase in $\Delta S$, the baryon asymmetry decreases (cf. Fig.~\ref{fig:m1-bound}). Thus, cogenesis of right DM abundance and baryon asymmetry is possible with an entropy production of the order of $\Delta S\lesssim\mathcal{O}(100)$.

In order to overcome the large entropy dilution at late epochs, we also check if the lepton asymmetry can be significantly overproduced. It is well established that the requirement on the heavy neutrino mass scale for successful leptogenesis can be significantly relaxed if (at least) two of the mass eigenvalues $M_i$, say for $i=1,2$ are quasi-degenerate, i.e., $\Delta M=M_2-M_1\ll \overline{M}=(M_1+M_2)/2$~\cite{Pilaftsis:2003gt, Dev:2017wwc}. In such a case of resonant leptogenesis, the CP-violating decay asymmetry originating from the interference of tree and self-energy contributions-the so-called $\epsilon$-type or indirect CP violation is {\it resonantly} enhanced, and dominates over the contribution from vertex-corrections, or the so-called $\epsilon'$-type or direct CP violation. Since a larger CP asymmetry results in larger baryon asymmetry, hence in this case one can allow larger entropy injection compared to the standard vanilla leptogenesis scenario. This in turn helps in relaxing the bound on DM mass, allowing heavier masses compared to the vanilla leptogenesis scenario. This is what we can see from the right panel of Fig.~\ref{fig:scan}, where we find the allowed parameter space corresponds to $\Delta S\gtrsim\mathcal{O}(100)$. Note that for $\Delta S\gtrsim 100$, the WDM limit becomes relatively relaxed and thus the whole parameter space opens up for correct DM relic. 

{
\section{Production from Gravity Mediated Scattering}
\label{sec:grav}
Apart from PBH, pure gravitational production of DM can also take place from the 2-to-2 scattering of the bath particles via $s$-channel mediation of massless graviton. The interaction rate density for such a process reads~\cite{Garny:2015sjg, Tang:2017hvq,Garny:2017kha, Bernal:2018qlk,Barman:2021ugy,Barman:2021qds}
\begin{equation}
    \gamma(T) = \alpha\, \frac{T^8}{M_\text{pl}^4}\,,
\end{equation}
with $\alpha \simeq 1.9\times 10^{-4}$ (real scalar), $\alpha \simeq 1.1\times 10^{-3}$ (Dirac fermion) or $\alpha \simeq 2.3\times 10^{-3}$ (vector boson). This kind of production is unavoidable due to universal coupling between the gravity and the stress-energy tensor involving the matter particles. The BEQ governing the time evolution of DM number density is thus given by
\begin{equation}\label{eq:beq-UV}
\dot n_\text{DM} + 3\,\mathcal{H}\,n_\text{DM}=\gamma\,.    
\end{equation}
\begin{figure}[htb!]
$$
\includegraphics[scale=0.34]{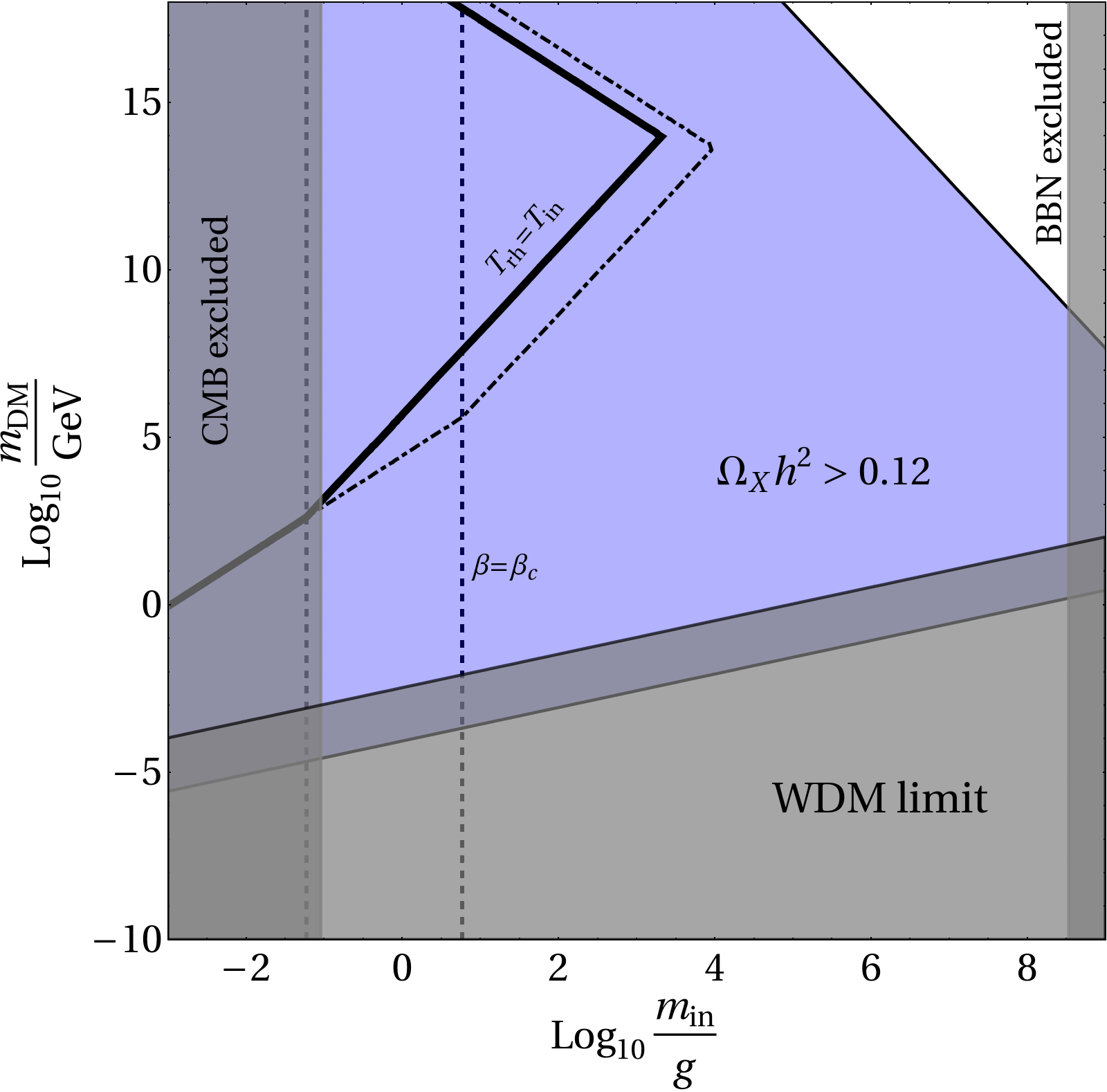}~~~~
\includegraphics[scale=0.34]{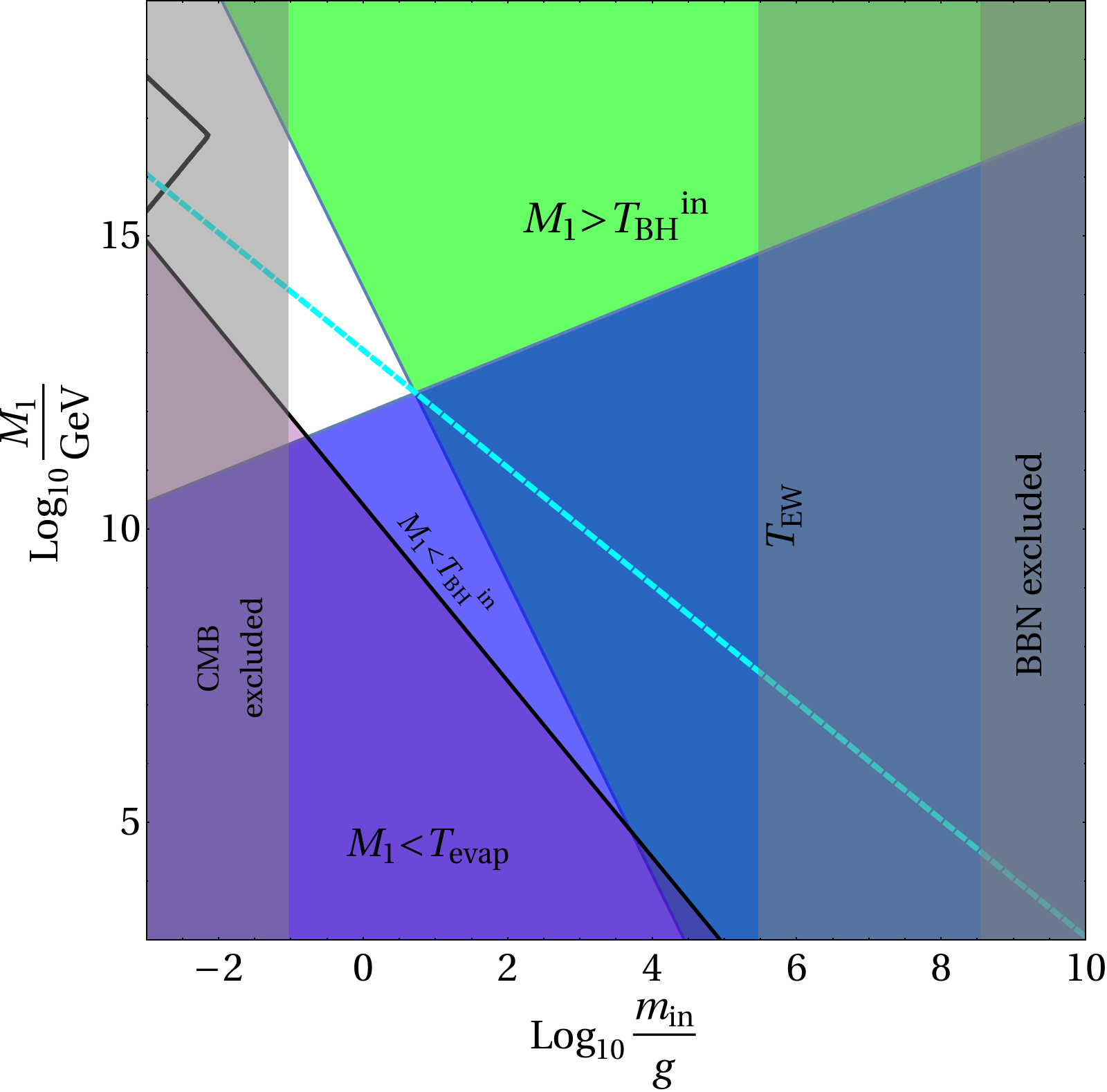}
$$
\caption{Left: The black thick line corresponds to right DM relic abundance via gravitational UV freeze-in. Along the contour we have $\Trh=T_\text{in}$. We consider no effect from entropy dilution (see text). Here the solid contour corresponds to $\beta=10^{-4}$ and the dashed one for $\beta=10^{-6}$. The straight vertical broken lines correspond to $\beta=\beta_c$. Right: The black solid and dashed contours correspond to observed baryon asymmetry for $\beta=10^{-4}$ and $\beta=10^{-6}$ respectively (the two contours overlap with each other and hence can not be distinguished). The coloured shaded regions are same as those in Fig.~\ref{fig:m1-bound}.}
\label{fig:grav-UV}
\end{figure}
For temperatures much lower than the reheat temperature i.e., $T\ll\Trh$, the DM yield can be analytically obtained by integrating Eq.~\eqref{eq:beq-UV}
\begin{equation}\label{eq:grav-yield1}
Y_0 = \frac{45\,\alpha}{2\,\pi^3\,g_{\star s}}\,\sqrt{\frac{10}{g_\star}}\,\left(\frac{T_\text{rh}}{M_\text{pl}}\right)^3\,,    
\end{equation}
where we define the DM yield as $Y\equiv n_\text{DM}/s$, with $s=\frac{2\,\pi^2}{45}\,g_{\star s}\,T^3$ and consider $\mdm\ll\Trh$. On the other hand, if the DM mass is such that $\Trh\ll \mdm\ll \Tmax$, where $\Tmax$ corresponds to the maximum temperature during reheating, then the DM can be produced during but not after the reheating. In the case the DM yield can be obtained by integrating Eq.~\eqref{eq:beq-UV} for $\Tmax\geq\mdm\geq\Trh$
\begin{equation}\label{eq:grav-yield2}
Y_0 =  \frac{45\,\alpha}{2\,\pi^3\,g_{\star s}}\,\sqrt{\frac{10}{g_\star}}\,\frac{\Trh^7}{M_\text{pl}^3\,\mdm^4}\,.     
\end{equation}
Here we would like to mention that if the DM is produced during the transition from matter to radiation domination via an interaction rate that scales like $\gamma(T)\propto T^n$, for $n>12$ the DM abundance is enhanced by a boost factor proportional to $(\Tmax/\Trh)^{n-12}$~\cite{Garcia:2017tuj}, whereas for $n\leq 12$ the results for the standard UV freeze-in calculation differ only by an $\mathcal{O}(1)$ factor from calculations taking into account of non-instantaneous reheating. 

Now, the DM produced via gravitational UV freeze-in shall undergo dilution due to evaporation of the PBH, that can be quantified as~\cite{Bernal:2021yyb,Bernal:2022oha} 
\begin{equation}
\frac{S(T_\text{in})}{S(T_\text{evap})}\simeq \frac{T_\text{evap}}{T_\text{peq}}\simeq 10^{-2} \left(\frac{M_\text{pl}}{\Min}\right)^\frac{3}{2}\,\frac{M_\text{pl}}{\beta\,T_\text{in}} \,,    
\end{equation}
for $\beta>\beta_c$, where we define $S=a^3\,s(T)$. The temperature $T_\text{peq}$ is defined as the epoch of equality between SM radiation and the PBH energy densities $\rho_R(T_\text{peq})=\rho_\text{BH}(T_\text{peq})$, and is given by
\begin{equation}
T_\text{peq} = \beta\,T_\text{in}\,\left(\frac{g_{\star,s}(T_\text{in})}{g_{\star,s}(T_\text{in})}\right)^\frac{1}{3}\,.    
\end{equation}
The observed DM abundance can then be achieved 
\begin{equation}
\mdm\,Y_0\,\frac{S(T_\text{in})}{S(T_\text{evap})}=\Omega_\text{DM}\,h^2\,\frac{1}{s_0}\,\frac{\rho_c}{h^2}\simeq 4.3\times 10^{-10}\,\rm GeV\,,   
\end{equation}
with $\rho_c$ being the critical density of the universe. In Fig.~\ref{fig:grav-UV} the black thick contour in the left panel satisfies correct DM relic abundance via gravitational UV freeze-in, considering $\Trh=T_\text{in}$. To the left of the contour, DM is over produced due to gravitational UV freeze-in. From this plot, it is clear that in the region of DM mass we are interested in (see Fig. \ref{fig:scan}), the DM production from the gravitational UV freeze-in remains under-abundant.

Similar to the case of DM, it is also possible to have leptogenesis from the decay of the RHNs, that are gravitationally produced from the SM bath via massless gravtion mediated scatterings~\cite{Bernal:2021kaj, Co:2022bgh}. Following the same methodology as above, in the right panel of Fig.~\ref{fig:grav-UV} we show the contour corresponding to right baryon asymmetry, using 
\begin{align}
& Y_B = \epsilon\,a_\text{sph}\,Y_B\Big|_{\Tev}\,,    
\end{align}
with $a_\text{sph}=28/79$ and $\epsilon$ is defined following Eq.~\eqref{eq:cp-asym}, where the RHN yield follows from Eq.~\eqref{eq:grav-yield1} and Eq.~\eqref{eq:grav-yield2}. Note that, the gravitational leptogenesis can dominate the production from PBH only for very light PBH which are already excluded from CMB bounds.
}

\section{Conclusions}
\label{sec:concl}
We have proposed a scenario where gravitational dark matter is produced from evaporating primordial black holes. Except for the superheavy mass window $m_\text{DM}\gtrsim 10^{10}$ GeV, DM production solely from PBH evaporation leads to overabundance if PBH dominates the energy density of the universe at early epochs. While lighter mass window $m_\text{DM}\lesssim \mathcal{O}(1)$ keV gives correct relic, it faces severe constraints from the requirement of structure formation. We particularly focus on this keV to $10^{10}$ GeV mass window of gravitational DM and incorporate it within a Type-I seesaw framework with three right handed neutrinos responsible for generating light neutrino masses. While DM in this mass window gets overproduced from PBH evaporation, late entropy injection from decay of one of the RHNs (acting as a diluter) can bring the DM abundance within observed limits. As the late entropy dilution must occur before the epoch of BBN in order not to disturb the successful prediction for light nuclei abundance, we impose the upper bound on diluter lifetime which gets translated into an upper bound on entropy injection, allowing DM mass upto $\sim 1$ PeV. Along with DM, the diluter also gets dominantly produced from PBH evaporation as its couplings to the SM particles remain suppressed from the requirement of long lifetime needed for sufficient entropy release due to its decay. This effectively leads to two different stages of early matter domination: first from PBH and then from the diluter. The other two RHNs can have sizeable couplings with SM leptons thereby generating the required neutrino mass and mixing. We consider the production of these RHNs both from the bath as well as PBH and show that their subsequent CP violating out-of-equilibrium decays can lead to successful leptogenesis as well. Since the lepton asymmetry is required to be overproduced initially in order to survive the subsequent entropy dilution, the DM parameter space gets squeezed from a few keV-PeV window to a smaller range around MeV-GeV ballpark from the requirement of producing both DM relic as well as the baryon asymmetry of the universe, with PBH mass $m_\text{BH}\lesssim 5$ g. The long-lived nature of $N_3$ necessarily pushes the lightest active neutrino mass $m_\nu^1$ to vanishingly small values ($\lesssim 10^{-10}$ eV). Thus, the effective neutrino mass will be very much out of reach from ongoing tritium beta decay experiments like KATRIN \cite{KATRIN:2019yun}, whereas any positive detection of $m_\nu^1$ in future experiments might be able to falsify our scenario. While gravitational DM has no scope of direct detection, the required PBH mass as well as multiple stages of early matter domination can have interesting observable consequences, specially in the context of gravitational wave observations~\cite{Anantua:2008am,Saito:2008jc, Hooper:2020evu, Papanikolaou:2020qtd, Kozaczuk:2021wcl, Bernal:2019lpc, Guo:2020grp, Borah:2022byb}, which we leave for future studies.

\appendix
\section{Coupled Boltzmann Equations}
\label{sec:cBEQ}

The relevant coupled Boltzmann equations (BEQ) in the present framework reads \cite{Perez-Gonzalez:2020vnz, JyotiDas:2021shi, Barman:2021ost} 
\begin{align}
&  \frac{dm_\text{BH}}{da} = -\frac{\kappa}{a\,\mathcal{H}}\,\epsilon(m_\text{BH})\,\left(\frac{1\text{g}}{m_{\rm BH}}\right)^2\,,
\nonumber\\&
\frac{d\widetilde{\rho}_R}{da}=-\frac{\epsilon_\text{SM}(m_\text{BH})}{\epsilon(m_\text{BH})}\,\frac{a}{m_\text{BH}}\,\frac{dm_\text{BH}}{da}\,\widetilde{\rho}_\text{BH} + \frac{a}{\mathcal{H}} \Gamma_{3} M_{3}\widetilde{n}_{N_3}^\text{BH}\,,
\nonumber\\& 
\frac{d\widetilde{\rho}_\text{BH}}{da}=\frac{1}{m_\text{BH}}\,\frac{dm_\text{BH}}{da}\,\widetilde{\rho}_\text{BH}\,,
\nonumber\\&
a\mathcal{H} \frac{d\widetilde{n}_{N_1}^T}{da}= -\left(\widetilde{n}_{N_1}^T-\widetilde{n}_{N_1}^\text{eq}\right)\,\Gamma^T_{N_1}\, ,
\nonumber\\&
a\mathcal{H}\frac{d\widetilde{n}_{N_1}^\text{BH}}{da}=-\widetilde{n}_{N_1}^\text{BH}\,\Gamma_{N_1}^\text{BH}+\Gamma_{\text{BH}\to N_1}\,\frac{\widetilde{\rho}_\text{BH}}{m_\text{BH}}\, ,
\end{align}
along with
\begin{align}
& a\mathcal{H}\frac{d\widetilde{N}_{B-L}}{da}=\epsilon_{\Delta L}\,\Biggl[\left(\widetilde{n}_{N_1}^T-\widetilde{n}_{N_1}^\text{eq}\right)\,\Gamma^T_{N_1}+\widetilde{n}_{N_1}^\text{BH}\,\Gamma_{N_1}^\text{BH}\Biggr]-\mathcal{W}\,\widetilde{N}_{B-L}\, ,
\nonumber\\&
a\mathcal{H}\frac{d\widetilde{n}_{N_3}^\text{BH}}{da}=\Gamma_{\text{BH}\to N_3}\,\frac{\widetilde{\rho}_{BH}}{m_{BH}} -\Gamma_{3}\widetilde{n}_{N_3}^\text{BH}
\nonumber\\&
a\mathcal{H}\frac{d\widetilde{n}_{DM}^\text{BH}}{da}=\Gamma_{\text{BH}\to DM}\,\frac{\widetilde{\rho}_{BH}}{m_{BH}}
\nonumber\\&
\frac{dT}{da}=-\frac{T}{\Delta}\Biggl[\frac{1}{a}+\frac{\epsilon_\text{SM}(m_\text{BH})}{\epsilon(m_\text{BH})}\,\frac{1}{m_\text{BH}}\,\frac{dm_\text{BH}}{da}\,\frac{g_\star(T)}{g_{\star s}(T)}\,a\,\frac{\widetilde{\rho}_{BH}}{4\,\widetilde{\rho}_{R}}+\frac{\Gamma_{3}M_{3}}{3 \mathcal{H} ~s~ a^4}\widetilde{n}_{N_3}^\text{BH}\Biggr]\,.    
\end{align}
Here, $\widetilde{\rho}_{R}$ and $\widetilde{\rho}_\text{BH}$ denotes the comoving energy density of radiation and PBH, $\mathcal{H}$ is the Hubble parameter, whereas $\widetilde{n}_{N1}$, $\widetilde{n}_{N3}$ and $\widetilde{n}_{DM}$ represent the comoving number densities of $N_1$, $N_3$ and DM respectively with $\epsilon_{\Delta L}$ being the CP-asymmetry parameter and $\mathcal{W}$ the term responsible for washout. Suffix `T' and `BH' denotes the thermal and non-thermal contributions respectively. Note that we are considering a scenario of DM with only gravitational interactions, hence it doesn't get produced from the thermal bath. Also, as we will see, $N_3$ plays the role of diluting the DM relic and hence have to be long-lived.  This results in suppressed Yukawas and hence no thermal production. Finally, $\widetilde{N}_{B-L}$ indicates the comoving B-L asymmetry. 

\section{Lyman-$\alpha$ Constraint}\label{sec:ly-alpha}
Due to their large initial momentum, DM particles could have a large free-streaming length leading to a suppression on the structure formation at small scales.
In the present scenario where DM has no interactions with the SM or with itself, the DM momentum simply redshifts, and its value $p_0$ at present is~\cite{Fujita:2014hha}
\begin{equation}
    p_0 = \frac{a_\text{evap}}{a_0}\, p_\text{evap}
    \simeq \frac{a_\text{evap}}{a_\text{eq}} \frac{\Omega_R}{\Omega_m}\, \langle E_\text{evap}\rangle = \langle E_\text{eq}\rangle\, \frac{\Omega_R}{\Omega_m}\,, 
\end{equation}
with
\begin{equation}
\langle E_\text{eq}\rangle = \langle E_\text{evap}\rangle\,\frac{a_\text{evap}}{a_\text{eq}}\approx \frac{\Tbh^\text{in}}{\xi}\,\frac{\Teq}{\Tev}\,\left[\frac{g_{\star,s}(\Teq)}{g_{\star,s}(\Tev)}\right]^\frac{1}{3}\,,     
\end{equation}
where in the last line we have assumed entropy injection from PBH evaporation { at $T=T_\text{evap}$} to matter-radiation equality { at $T=T_\text{eq}$} as $\xi\,\left(sa^3\right)_\text{evap}=\left(sa^3\right)_\text{eq}$~\cite{Masina:2020xhk}. { Note that the average kinetic energy of the emitted particles depend on the Hawking temperature, and is given by $\langle E_\text{evap}\rangle=6\,\Tbh^\text{in}$~\cite{Fujita:2014hha}. However, a more refined calculation shows the factor 6 to be approximately 1.3~\cite{Masina:2020xhk}}. A lower bound on the DM mass can be obtained from the upper bound on a typical velocity of warm DM at the present time. Taking $v_\text{DM} \lesssim 1.8 \times 10^{-8}$~\cite{Masina:2020xhk} for $\mdm \simeq 3.5$~keV~\cite{Irsic:2017ixq}, we obtain
\begin{align}\label{eq:wdm-limit}
&  \mdm \gtrsim  10^4\,\langle E_\text{eq}\rangle\equiv 10^4\,\frac{\Teq}{\xi}\,\frac{\Tbh^\text{in}}{\Tev}\,\left[\frac{g_{\star,s}(\Teq)}{g_{\star,s}(\Tev)}\right]^\frac{1}{3}
\simeq 2\times 10^4\,\frac{\Teq}{\xi}\,\sqrt{\frac{m_\text{BH}^\text{in}}{M_\text{pl}}}\,\left[\frac{g_{\star,s}(\Teq)}{g_{\star,s}(\Tev)}\right]^\frac{1}{3}\,,
\end{align}
where $a_\text{eq}\equiv\frac{\Omega_R}{\Omega_m}\simeq 1.8\times 10^{-4}$ and $\Teq\simeq 0.75$ eV.

\section{Casas-Ibarra Parametrization}\label{sec:app-CI}

As the neutral component of the SM Higgs doublet acquires a VEV leading to the spontaneous breaking of the SM gauge symmetry, neutrinos in the SM obtain a Dirac mass that can be written as
\begin{align}
m_D= \frac{y_{N}}{\sqrt{2}}v.
\end{align}
The Dirac mass $m_D$ together with the RHN bare mass $M_N$, can explain the nonzero light neutrino masses with the help of Type-I seesaw~\cite{GellMann:1980vs, Mohapatra:1979ia} where the light-neutrino masses can be expressed as,
\begin{align}
m_{\nu}\simeq m_{D}^T~M^{-1}~m_{D}.
\end{align}

The mass eigenvalues and mixing are then obtained by diagonalising the light-neutrino mass matrix as
\begin{align}
m_{\nu}=\mathcal{U}^* m_{\nu}^d \mathcal{U}^{\dagger}\,,
\end{align}
with $m_{\nu}^d=dia(m_1,m_2,m_3)$ consisting of the mass eigenvalues and $\mathcal{U}$ representing the Pontecorvo-Maki-Nakagawa-Sakata matrix~\cite{Zyla:2020zbs}\footnote{The charged lepton mass matrix is considered to be diagonal.}. 
One can parameterise the complex structure of the neutrino Yukawa matrix that is required for generating the CP violating decays of the RHNS using Casas-Ibarra (CI) parametrisation~\cite{Casas:2001sr} as,
\begin{align}
y_N = \frac{\sqrt{2}}{v}\sqrt{M}~\mathbb{R}~\sqrt{m_{\nu}^d}~\mathcal{U}^{\dagger}\,,
\label{CI}
\end{align}
where $\mathbb{R}$ is a complex orthogonal matrix $\mathbb{R}^T \mathbb{R} = I$, which we choose as
\begin{align}
\mathbb{R} =
\begin{pmatrix}
0 & \cos{z} & \sin{z}\\
0 & -\sin{z} & \cos{z}\\
1 & 0 & 0
\end{pmatrix}\,,
\label{eq:rot-mat}
\end{align} 
where $z=a+ib$ is a complex angle.  
The diagonal light neutrino mass matrix $m_{\nu}^d$ is calculable using the best fit values of solar and atmospheric mass  obtained from the latest neutrino oscillation data~\cite{Zyla:2020zbs}. Now, the elements of Yukawa coupling matrix $y_N$ for a specific value of $z$, can be obtained for different choices of the heavy neutrino masses. For example, with $M_1=10^{13}$ GeV, $M_2=50\,M_1$, $M_3=10^{9}$ GeV, $m_\nu^1=10^{-15}$ eV and  $\{a,b\}=\{2.77,0.36\}$ we obtain the following structure
\begin{align}
y_N=\left(
\begin{array}{ccc}
 -0.037\, - 0.007\,i& -0.002\, - 0.031\,i &  0.064\, - 0.026\,i \\
 -0.033\, + 0.197\,i & -0.682\, + 0.0126\,i & -0.592\, - 0.152\,i \\
 1.508 \times 10^{-10}& -6.764 \times 10^{-11} - 1.313\times10^{-11} i & 7.30 \times 10^{-11} - 1.490 \times 10^{-11}\,i
\end{array}
\right)\,,
\end{align}
which satisfies the light neutrino mass, as well as produces desired CP asymmetry.

\section*{Acknowledgements}
BB received funding from the Patrimonio Autónomo - Fondo Nacional de Financiamiento para la Ciencia, la Tecnología y la Innovación Francisco José de Caldas (MinCiencias - Colombia) grant 80740-465-2020. This project has received funding /support from the European Union's Horizon 2020 research and innovation programme under the Marie Sklodowska-Curie grant agreement No 860881-HIDDeN. RR was supported by the National Research Foundation of Korea (NRF) grant funded by the Korean government (NRF-2020R1C1C1012452). 

\bibliographystyle{JHEP}
\bibliography{Bibliography}

\end{document}